\documentclass{WileyMSP-template}

\begin{document}

\newcommand{\czts}{Cu$_2$ZnSnS$_4$}
\newcommand{\cztsse}{Cu$_2$ZnSn(S,Se)$_4$}
\newcommand{\czgs}{Cu$_2$ZnGeS$_4$}
\newcommand{\cigs}{CuInGa(S,Se)$_2$}
\newcommand{\etc}{\textit{etc.}}
\newcommand{\cfr}{\textit{cfr.}}
\newcommand{\etal}{\textit{et al.}}
\newcommand{\voc}{$V_{\mathrm{OC}}$}
\newcommand{\jsc}{$J_{\mathrm{SC}}$}
\newcommand{\ai}{\textit{ab initio}}
\newcommand{\ie}{i.\,e.,}
\newcommand{\angstrom}{\mbox{\normalfont\AA}}

\pagestyle{fancy}

\title{Relevance of Ge incorporation to control the physical behaviour of point defects in kesterite}

\maketitle

% Author: Please give full first and last names for authors and include * after the name of all corresponding authors

\author{Thomas Ratz*}
\author{Ngoc Duy Nguyen}
\author{Guy Brammertz}
\author{Bart Vermang}
\author{Jean-Yves Raty}

% Dedication

%\dedication{Optional dedication here. If no dedication is required, please leave blank}

% Affiliations: Please provide adacemic titles (Prof. or Dr.) for all authors where applicable, and include an institutional email address for all corresponding authors
\begin{affiliations}
Thomas Ratz
CESAM | Q-MAT | Solid State Physics, Interfaces and Nanostructures, Physics Institute B5a, Allée du Six Août 19, B-4000 Liège, Belgium \\
Institute for Material Research (IMO), Hasselt University, Agoralaan gebouw H, B-3590 Diepenbeek, Belgium\\
thomas.ratz@uliege.be \\

Prof. Ngoc Duy Nguyen \\
CESAM | Q-MAT | Solid State Physics, Interfaces and Nanostructures, Physics Institute B5a, Allée du Six Août 19, B-4000 Liège, Belgium \\

Dr. Guy Brammertz \\
IMEC division IMOMEC | partner in Solliance, Wetenschapspark 1, B-3590 Diepenbeek, Belgium \\

Prof. Bart Vermang \\
Institute for Material Research (IMO), Hasselt University, Agoralaan gebouw H, B-3590 Diepenbeek, Belgium \\
IMEC division IMOMEC | partner in Solliance, Wetenschapspark 1, B-3590 Diepenbeek, Belgium\\
Energyville, Thor Park 8320, B-3600 Genk, Belgium\\

Dr. Jean-Yves Raty \\
CESAM | Q-MAT | Solid State Physics, Interfaces and Nanostructures, Physics Institute B5a, Allée du Six Août 19, B-4000 Liège, Belgium
\end{affiliations}

% Keywords: Please provide a minimum of three and a maximum of seven keywords, separated by commas

\keywords{kesterite, first-principles calculations, point defect, Ge doping and alloying}

% Abstract should be written in the present tense and impersonal style (i.e., avoid we), and be at most 200 words long
\begin{abstract}

To reduce the prominent $V_{OC}$-deficit that limits kesterite-based solar cells efficiencies, Ge has been proposed over the recent years with encouraging results, as the reduction of the non-radiative recombination rate is considered as a way to improve the well-known Sn-kesterite world record efficiency. To gain further insight into this mechanism, we investigate the physical behaviour of intrinsic point defects both upon Ge doping and alloying of \czts \ kesterite. Using a first-principles approach, we confirm the p-type conductivity of both \czts \ and \czgs , attributed to the low formation energies of the $\mathrm{V_{Cu}}$ and $\mathrm{Cu_{Zn}}$ acceptor defects within the whole stable phase diagram range. Via doping of the Sn-kesterite matrix, we report the lowest formation energy for the substitutional defect $\mathrm{Ge_{Sn}}$. We also confirm the detrimental role of the substitutional defects $\mathrm{X_{Zn}}$ (X=Sn,Ge)  acting as recombination centres within the Sn-based, the Ge-doped and the Ge-based kesterite. Finally, we highlight the reduction of the lattice distortion upon Ge incorporation resulting in a reduction of the carrier capture cross section and consequently a decrease of the non-radiative recombination rate within the bulk material.

\end{abstract}

% Text: Please use section headings and subheadings as specified below. For communications, all section headings apart from Experimental Section should be removed
% Please make the first reference to a display item bold: \textbf{Figure 1}
% Do not abbreviate Figure, Equation, etc.; display items are always singular, i.e., Figure 1 and 2.
% Equations are always singular, i.e., Equation 1 and 2, and should be inserted using the {equation} environment, not as graphics
% Please do not use footnotes in the text, additional information can be added to the Reference list.

%///////////////////////////////////////////////////////////////////////////////////////////////////////////////////////////////////////////////////////////////////////////////////////////////////////////
\section{Introduction}
\label{Introduction}

Emerging solar cell technologies often struggle with minority carrier lifetime, leading to large open circuit voltage deficit ($V_{OC}$-deficit) and consequently low solar cell efficiency. With no exception, in opposition to its well-established CuInGa(S,Se)$_2$ chalcogenide parent showing high cell efficiency \cite{Green:2020ga, nakamura2019cd}, kesterite materials are dealing with efficiency issues \cite{Giraldo:2019iia, Ratz:2019cs, todorov2020solution}. As possible culprits of the actual limitation, electronic defects acting as recombination centres have been pointed out \cite{Chen:2013cna,Kim:2018jd, grossberg2019electrical, pal2019current, dimitrievska2016secondary, schorr2019point} together with other obstacles such as band alignments \cite{PlatzerBjorkman:2019ed, crovetto2017band, pal2019current}, secondary phases \cite{dimitrievska2016secondary, schorr2019point} and/or band tailing caused by electrostatic potential fluctuations due to the presence of charged electronic defects \cite{gokmen2013band, grossberg2019electrical, pal2019current}.

Focusing on defects, as a result of the complex structure of kesterite materials, a wide range of intrinsic and cluster defects can form within the crystal, leading to various impacts on the kesterite absorber layer opto-electronic properties. Using first-principles calculations, Chen \etal \ were able to predict the p-type conductivity of Cu$_2$ZnSn(S,Se)$_4$ via the high population of the $\mathrm{Cu_{Zn}}$ and $\mathrm{V_{Cu}}$ defects while identifying the [2 $\mathrm{Cu_{Zn}} + \mathrm{Sn_{Zn}}$] cluster defect as recombination centre leading to charge carrier losses \cite{Chen:2013cna}. Experimentally, Dimitrievska \etal \ reported the possible tuning of the $V_{OC}$ value according to the Cu concentration and consequently the amount of [$\mathrm{V_{Cu}}$ + $\mathrm{Zn_{Cu}}$] defect clusters \cite{dimitrievska2016secondary}. More recently, Kim \etal \ identified the intrisinc defect $\mathrm{Sn_{Zn}}$ as the origin of the electron capture and emission in the \czts \ compound resulting of the Sn multivalence \cite{Kim:2018jd, biswas2010electronic, gong2021identifying}. Furthermore, Gong \etal \ established a link between the Sn oxidation states Sn$^{+2}$ and Sn$^{+4}$ and the kesterite growth conditions \cite{gong2021identifying}. Consequently, gaining further knowledge into point defects and cluster defects in kesterite materials could allow the control of the absorber layer physical properties in view of increasing the kesterite-based cells efficiencies.

Over the recent years, attempts have been made to circumvent these actual limitations using alloying and doping of kesterite materials with other elements \cite{Romanyuk:2019cq, crovetto2020assessing, Jyothirmai:2019gb, Giraldo:2019iia, li2018cation,neuschitzer2018revealing}. Both theoretical and experimental approaches have been used. A wide range of cationic substitutions have been investigated: Cu by Ag \cite{yuan2015engineering}, Zn by Cd \cite{sharif2020control, yuan2015engineering}, Sn by Ge \cite{khelifi2021path,vermang2019wide, choubrac2020sn, giraldo2015large}, S by Se \cite{ratz2019physical} and doping of both \czts \ and Cu$_2$ZnSnSe$_4$ by Na, Li, Ga \cite{du2021defect} and Ge \cite{kim2016improvement, giraldo2018small, deng2021adjusting,Romanyuk:2019cq} or even using more exotic elements as in Ref.\cite{Jyothirmai:2019gb}. Some of these substitutions resulted in cell efficiencies as large as 12.3 \% as in the case of Ga or Ge doping \cite{du2021defect, kim2016improvement}. Double cation incorporation like Ge and Ag \cite{fu2020ag} or Ge and Cd  \cite{he2021systematic} were also realised simultaneously allying the benefits of both substitutional elements. Nevertheless, the reported efficiencies are still below the world-record of pure Sn-kesterite compound \cite{wang2014device}. 

From these investigations, Ge emerged as an interesting doping/alloying element as several studies reported high solar cell efficiencies through the improvement of the $V_{OC}$ values following the incorporation of small amounts of Ge \cite{giraldo2018small} or via the complete substitution of Sn by Ge \cite{choubrac2020sn}. In a recent study, Deng \etal \ demonstrated experimentally that Ge$^{4+}$ can be introduced in \czts \ to suppress the detrimental deep $\mathrm{Sn_{Zn}}$ defects \cite{deng2021adjusting}. In addition, compared to \czts , pure Ge-kesterite absorber layers present a larger band gap value which limits the maximal single solar cell efficiency \cite{ratz2021opto}. Nevertheless, Ge alloying could be used for the synthesis of wide band gap kesterite which is of high interest for top cells in a tandem approach \cite{khelifi2021path,vermang2019wide}. Gaining further insight into the physical behaviour of intrinsic defects in Ge-kesterite and into the mechanisms of Ge doping within the Sn-kesterite is therefore strongly desirable.

Moreover, although several material modelling studies have focused on the physical behaviour of defects in kesterite compounds \cite{Chen:2013cna, Kim:2018jd, Jyothirmai:2019gb, du2021defect}, only a few works have been dedicated so far to Ge compounds \cite{nishihara2017first}. To fill this void, in this work, we report the investigation of point defects in both \czts \ and \czgs . We first study the physical behaviour of intrinsic point defects in both kesterite materials to highlight the impact of Ge alloying and secondly, we investigate Ge-related point defects in the Sn-kesterite compound to illustrate the Ge doping mechanism. Focusing on the physical behaviour of point defects in these materials, our aim is to establish a link between the growth conditions of the kesterite thin films, the formation of point defects and the resulting kesterite solar cell performances by identifying the defect physical behaviour (dopant or recombination centre). In addition, based on the empirical rule proposed by Li \etal \ we derive meaningful trends concerning the defect carrier capture cross section. Indeed, it was reported that this quantity can be related to the lattice distortion caused by the defect incorporation in its various charge states \cite{li2019effective}.

The paper is organised as follows. The phase diagrams of the two kesterites are first presented. Once proper chemical potential ranges are set, avoiding secondary phases and obtaining the desired kesterite phase, the defect formation energies are obtained according to the Fermi level position. Then, we present the defect charged states and their possible ionisation levels in order to evaluate their physical behaviour in their various electronic configurations. This approach allows us to identify their roles as dopants or as recombination centre. Finally, in the last part of this paper, we present a study of the lattice distortions around the different incorporated defects. This information allows us to provide a guide for the extraction of general trends concerning the capture cross sections of the various point defects at play. As a result, we are able to predict which point defects are the most abundant in the kesterite materials and eventually, to characterise the defect levels (acceptor, donor or recombination centre) and their relative impact on the capture of charge carriers. Considering various locations in the phase diagram, we also consider different growth conditions of the kesterite materials. This study allows to highlight (i) the impact of Ge-alloying of kesterite on the intrinsic point defects physical behaviour and (ii) the physical behaviour of Ge dopants in \czts .

\section{Theoretical framework}
\label{theoreticalframework}

\begin{figure}[H]
\centering
\includegraphics[width=0.4\textwidth]{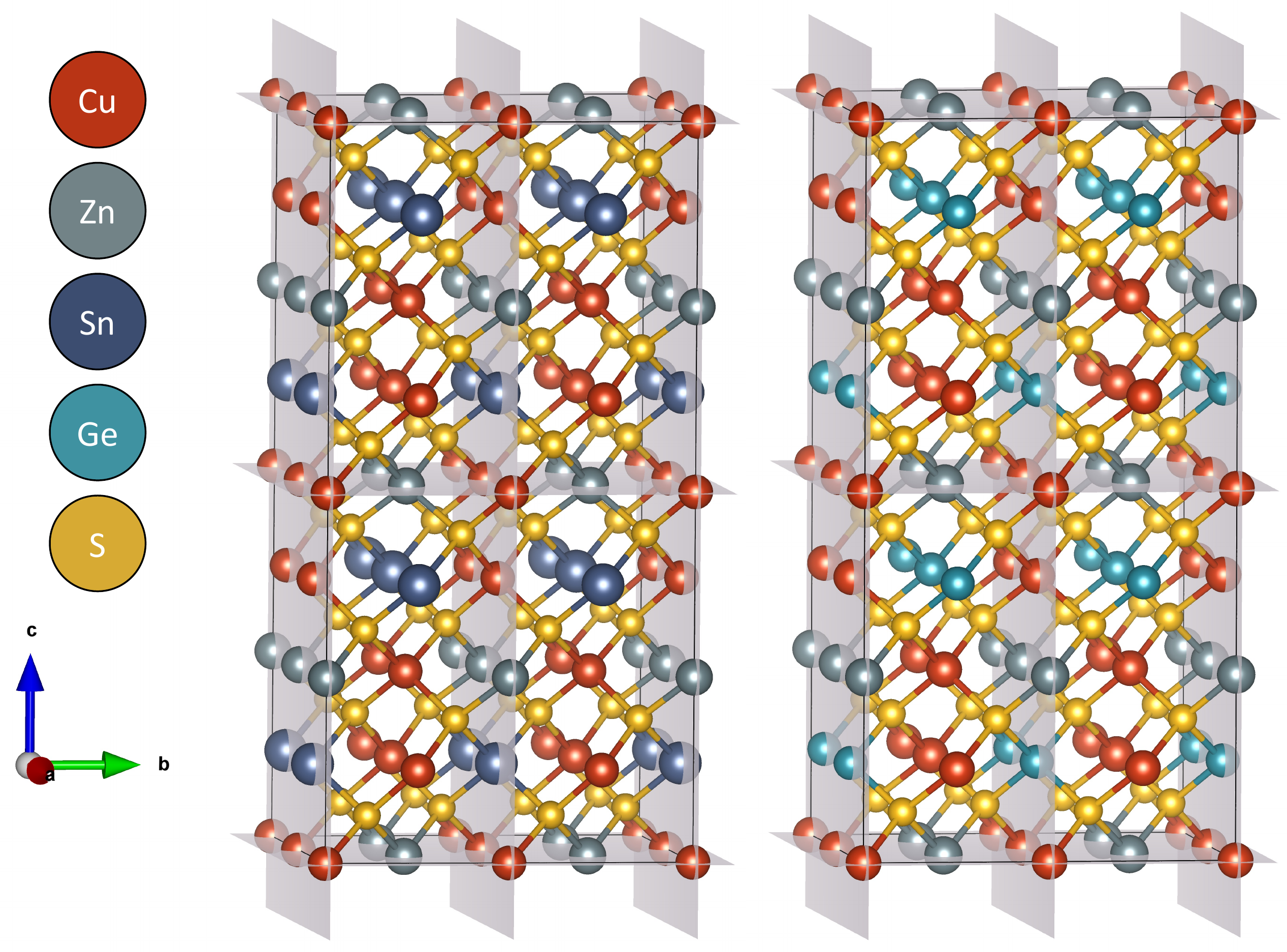}
\caption{Pristine kesterite 64-atoms supercells used to compute the defect formation energies. Each supercell corresponds to an expansion of $2 \times 2 \times 2$ of the kesterite conventional cell as represented by the grey shadings.} 
\label{Kesterite_structures}
\end{figure}

Beyond materials properties predictions, the first-principles approach is a powerful tool to understand the behaviour of defects in semiconductor compounds \cite{freysoldt2014first, lany2008assessment, persson2005n, park2018point}. In this work, the supercell approach is considered and the calculations are performed with a 64-atoms supercell corresponding to an expansion of $2 \times 2 \times 2$ of the kesterite conventional cell as presented in \textbf{Figure \ref{Kesterite_structures}}. 

Using this approach, the formation energy of a defect $\alpha$ in a charge state $q$ can be calculated as follows,

\begin{equation}
\Delta H_F (\alpha,q, E_F, \mu_i) = E (\alpha,q) - E_{\mathrm{host}} - \sum_i n_i (E_i + \mu_i ) + q [\epsilon_{\mathrm{VBM,host}}+ E_F],
\label{Formation_energy_eq}
\end{equation}
where $E (\alpha,q)$ is the total energy of the supercell with a defect $\alpha$ in the charge state $q$; $E_{\mathrm{host}}$ is the total energy of the 64-atom pristine supercell. $n_i$ is the number of atom(s) of the species $i$ removed ($<0$) from or added ($>0$) to the host supercell with $E_i$, the energy per atom of the pure phase of the species $i$ and $\mu_i$, the chemical potential of the corresponding element. A variation in the synthesis conditions can change the thin film composition and, consequently, the environment in which the defect will be formed. As a result, the point defect formation energy and the defect concentration will be impacted according the amount of energy required by the exchange of particles necessary to form the defect. This energy cost is described by the chemical potential of the chemical species ($\mu_i$) which is defined as the Gibbs free energy variation caused by the exchange of particles between the system and an external reservoir, $\mu_i = (\frac{dG}{dN_i})$ \cite{freysoldt2014first}. To obtain the chemical potentials values $\mu_i$ leading to the formation of a stable kesterite phase without secondary phases, the kesterite phase diagrams have first to be computed. To do so, a set of thermodynamic conditions must be fulfilled as presented in the supplementary information (SI). Then, assuming a defect with a charge state $q$, a fourth term is added to Equation \ref{Formation_energy_eq}. In this term, $\epsilon_{\mathrm{VBM,host}}$ refers to the valence band maximum (VBM) of the host supercell and $E_F$ is the Fermi level acting as a parameter of the defect formation energy function and ranging from the VBM to the band gap energy $E_G$ of the kesterite material. Furthermore, based on the formation energy of a given defect, the position of its ionisation levels in the materials band gap can be obtained using the following relation:

\begin{equation}
\mathcal{E}(\alpha, q,q') = \frac{\Delta H_F(\alpha , q) - \Delta H_F(\alpha ,q')}{q' - q},
\label{ionization_eq}
\end{equation}
with $q,q'$, the charge states of the defect $\alpha$ considered for the transition.

\section{Results and discussion}
\label{resultanddiscussion}

\subsection{Kesterite growth: chemical environment}

\begin{figure}[H]
\subfloat[Cu$_2$ZnSnS$_4$ - $\mu_{\mathrm{Cu}} = -0.27$ eV]{\includegraphics[width=0.33\textwidth]{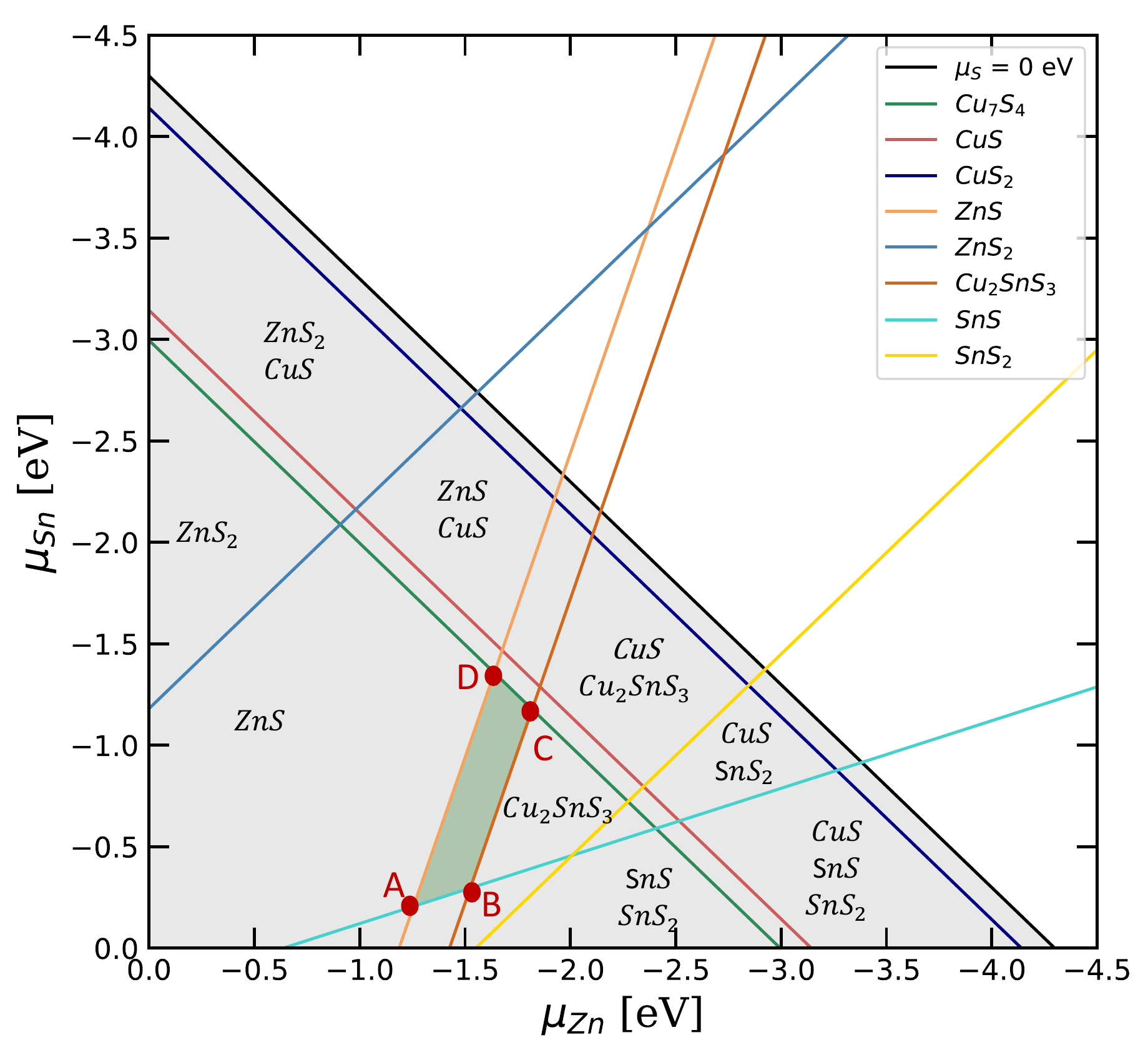}\label{PD_CZTS_1}} \hfill
\subfloat[Cu$_2$ZnSnS$_4$ - $\mu_{\mathrm{Cu}} = -0.55$ eV]{\includegraphics[width=0.33\textwidth]{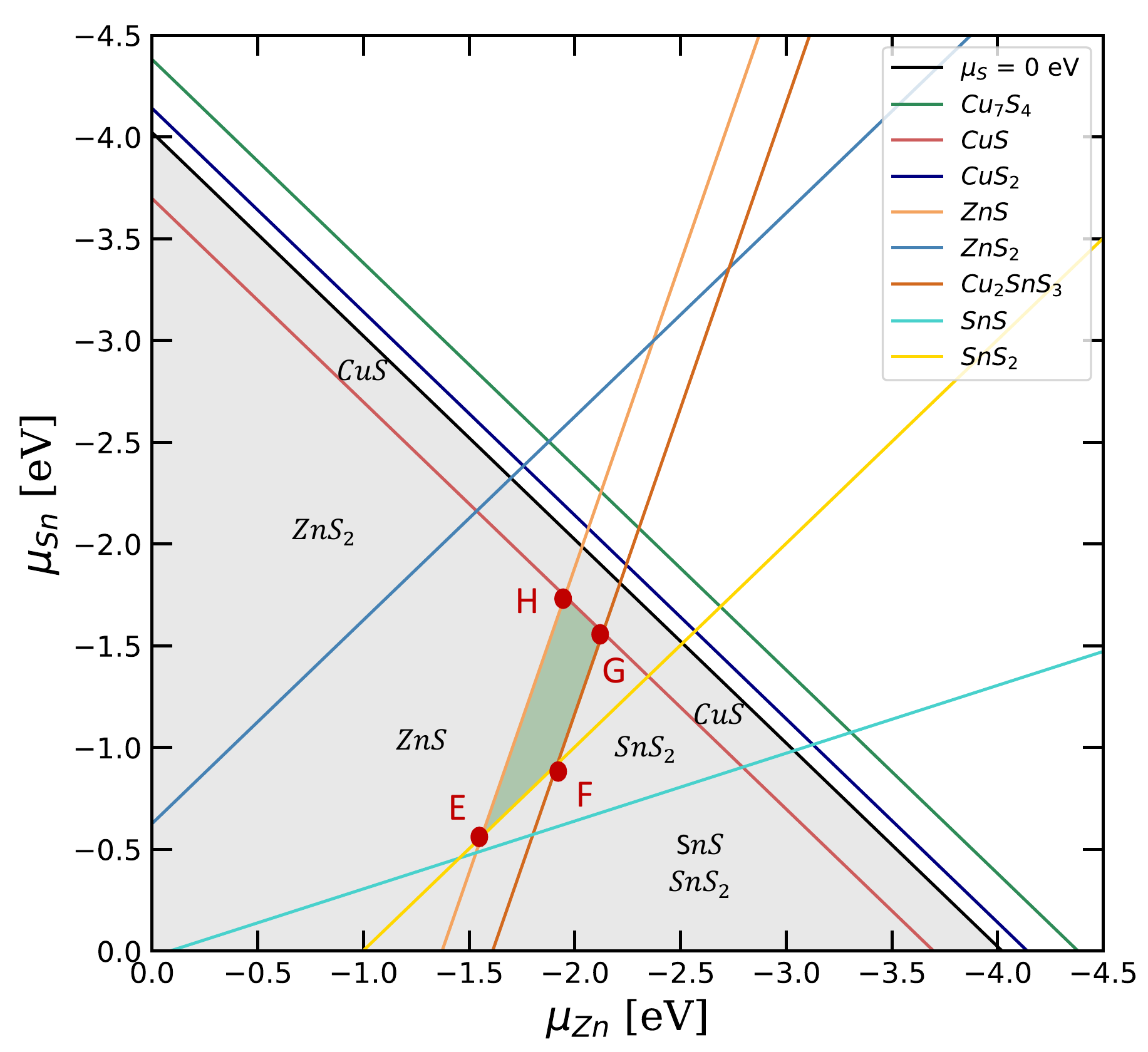}\label{PD_CZTS_2}} \hfill
\subfloat[Cu$_2$ZnSnS$_4$ - $\mu_{\mathrm{Cu}} = -0.82$ eV]{\includegraphics[width=0.33\textwidth]{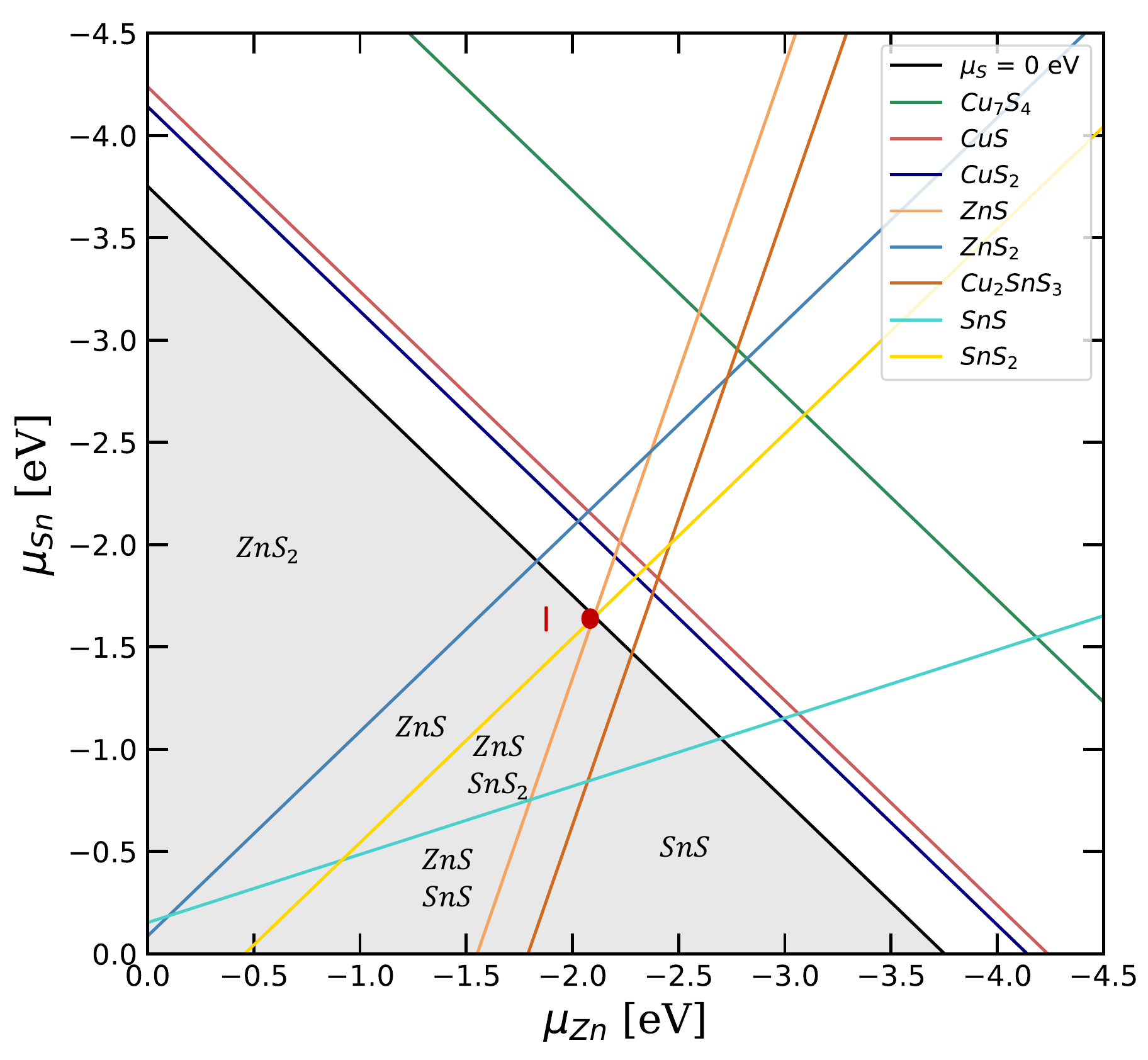}\label{PD_CZTS_3}} \\
\subfloat[Cu$_2$ZnGeS$_4$ - $\mu_{\mathrm{Cu}} = -0.27$ eV]{\includegraphics[width=0.33\textwidth]{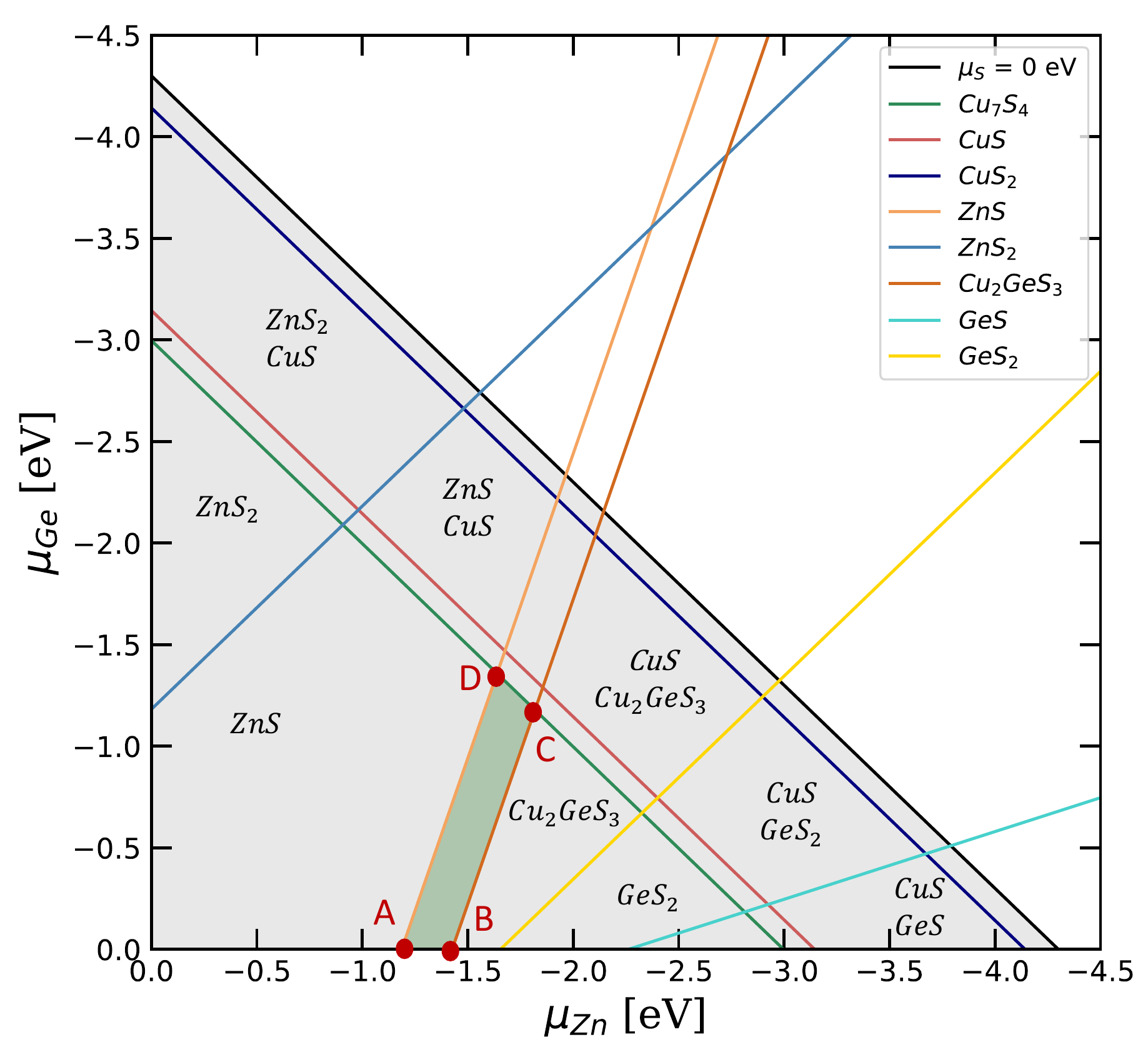}\label{PD_CZGS_1}} \hfill
\subfloat[Cu$_2$ZnGeS$_4$ - $\mu_{\mathrm{Cu}} = -0.55$ eV]{\includegraphics[width=0.33\textwidth]{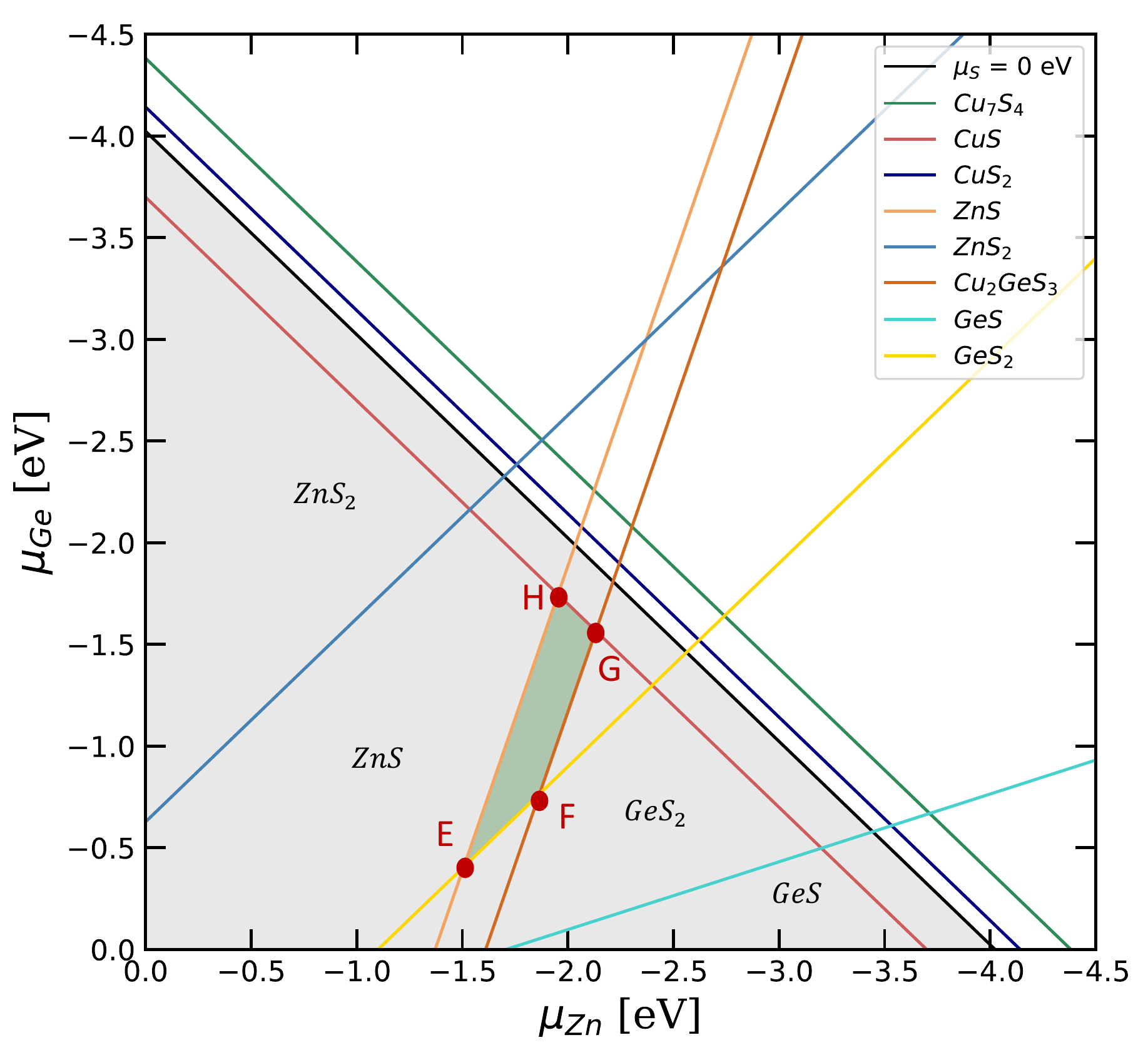}\label{PD_CZGS_2}} \hfill
\subfloat[Cu$_2$ZnGeS$_4$ - $\mu_{\mathrm{Cu}} = -0.82$ eV]{\includegraphics[width=0.33\textwidth]{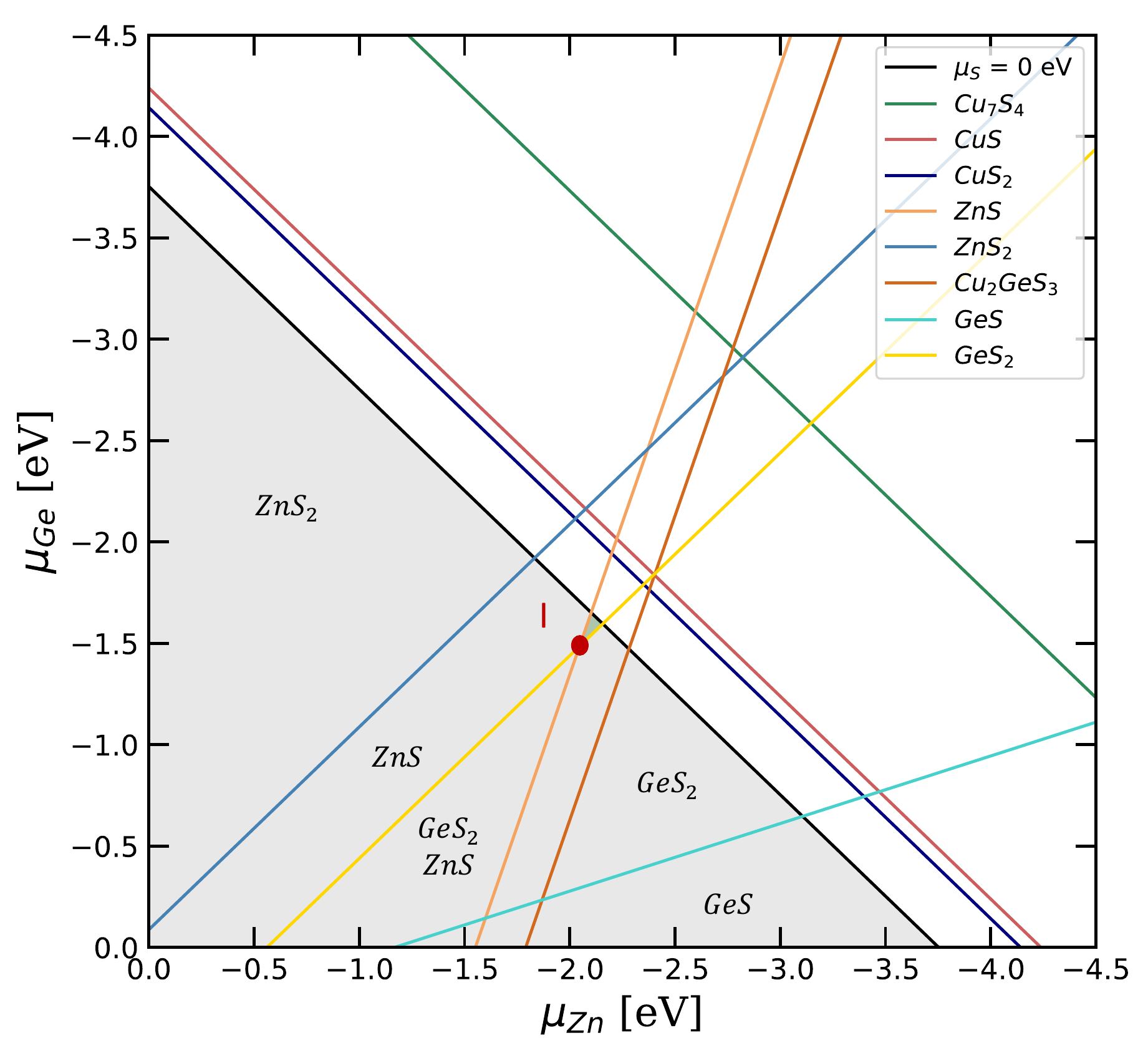}\label{PD_CZGS_3}}
\caption{Phase diagrams of \czts \ and \czgs for three different Cu concentrations. Each line corresponds to a secondary phase as listed in the legends (see SI for secondary phases calculations). The grey-shaded areas correspond to chemical potential values for which the kesterite can be thermodynamically synthesised (stable phase). Upon taking into account the possible presence of the secondary phases, the green shading indicates the chemical potential ranges for which a kesterite stable phase is encountered without any secondary phase.}
\label{PD}
\end{figure}

Let us first determine the chemical environment in which the kesterite materials can be synthesised without secondary phases. In \textbf{Figure \ref{PD}}, the phase diagrams of Sn-based and Ge-based kesterites are presented for three different Cu concentrations: $\mu_{Cu}$ = -0.27 (Cu-rich) eV, $\mu_{Cu}$ = -0.55 eV (Cu-moderate) and $\mu_{Cu}$ = -0.82 eV (Cu-poor). 

A first general observation is the narrow chemical potential range in which the kesterite phase can be found without any secondary phase (green shading) in comparison to the possible chemical potentials combinations that could lead to the formation of the kesterite (grey shading). For lower Cu concentrations, corresponding to a $\mu_{Cu}$ value lower than -0.82 eV (resp. -0.85 eV for \czgs ), no pure kesterite phase is available within stable combinations of chemical potentials values. This means that below such a critical $\mu_{Cu}$ value, the kesterite phase cannot be formed without the presence of secondary phases. In addition, shifting from Figure \ref{PD_CZTS_1} to Figure \ref{PD_CZTS_3}, we observe a reduction of the stable kesterite phase area (grey shading). This reduction occurs as we move towards lower Cu concentration (see supplementary information). If one chemical potential absolute value increases for a constant kesterite formation energy, the available chemical potential combinations range is reduced (\textit{i.e.} the edge of the stable kesterite phase corresponding to $\mu_S = 0$ is shrunk). Physically, it means that for a lower chemical potential value corresponding to a lower concentration, another element concentration has to increase to keep obtaining a stable kesterite phase.

The reduction of the stable kesterite areas (grey shading) to the pure kesterite areas (green shading) is the result of the multiple secondary phases that can be formed using the 4 chemical elements composing the kesterite compounds. As shown in Figure \ref{PD_CZTS_1}, around the green area, the dominant secondary phases are ZnS, Cu$_7$S$_4$, Cu$_2$SnS$_3$ and SnS. The phases predicted by our calculations are in good agreement with those reported by Chen \etal \ in Ref.\cite{Chen:2013cna}. In addition, in the recent review of Schorr \etal , the authors reported the presence of ZnS in each thin film where the composition Zn/Sn exceeded the value of 1. It was also reported that the concentrations of both ZnS and SnS secondary phases increase upon deviation from the stoichiometry \cite{schorr2019point, just2016secondary, dimitrievska2016secondary}. Similarly, for the Ge-compound in Figure \ref{PD_CZGS_1}, the secondary phases located next to the pure kesterite phase are ZnS, Cu$_7$S$_4$ and Cu$_2$GeS$_3$. Experimentally, using XRD and HAADF-STEM imaging, Khelifi \etal \ held the ZnSe secondary phase at the top of the absorber layer and a thick Cu$_2$GeSe$_3$ secondary phase (120-160 nm) at the bottom of the kesterite thin film accountable for the main efficiency loss \cite{khelifi2021path}. In the case of the Ge-based compound and in comparison with the SnS phase in the Sn-kesterite, the GeS secondary phase appears to be shifted towards positive Ge chemical potential values. As a result, the Ge-kesterite pure phase area (green shading) is slightly larger compared to the Sn-kesterite one. Following the phase diagram evolution from Figure \ref{PD_CZTS_1} to Figure \ref{PD_CZTS_3}, we observe that, as the absolute value of the Cu chemical potential increases, the SnS$_2$ and SnS secondary phase limits shift towards lower Sn concentrations while the ZnS secondary edge shifts towards lower Zn concentrations until both  SnS$_2$ and ZnS secondary phase edges cross each other for a Cu chemical potential value of $\mu_{Cu}$=-0.82 eV, leading to the vanishing of the pure kesterite area. The same behaviour can be observed for \czgs \ with an extreme copper chemical potential value of -0.85 eV involving the following secondary phases: GeS, GeS$_2$ and ZnS. Recently, Choubrac \etal \ reported for \czgs \ the effectiveness of a set of different surface treatments to get rid off all detrimental secondary phases \cite{choubrac2020sn}. 

Then, to compute Ge-doping point defect in the Sn-based kesterite, one has to obtain a chemical potential value for the Ge ($\mu_{Ge}$). To do so, as presented in the SI, several additional secondary phases were computed. 

In the next section, we will focus on the formation energy of the point defects using chemical potential combinations providing a pure kesterite phase.

\subsection{Intrinsic point defects formation energies}

\begin{figure*}[t]
\centering
\subfloat[Cu$_2$ZnSnS$_4$]{\includegraphics[width=0.55\textwidth]{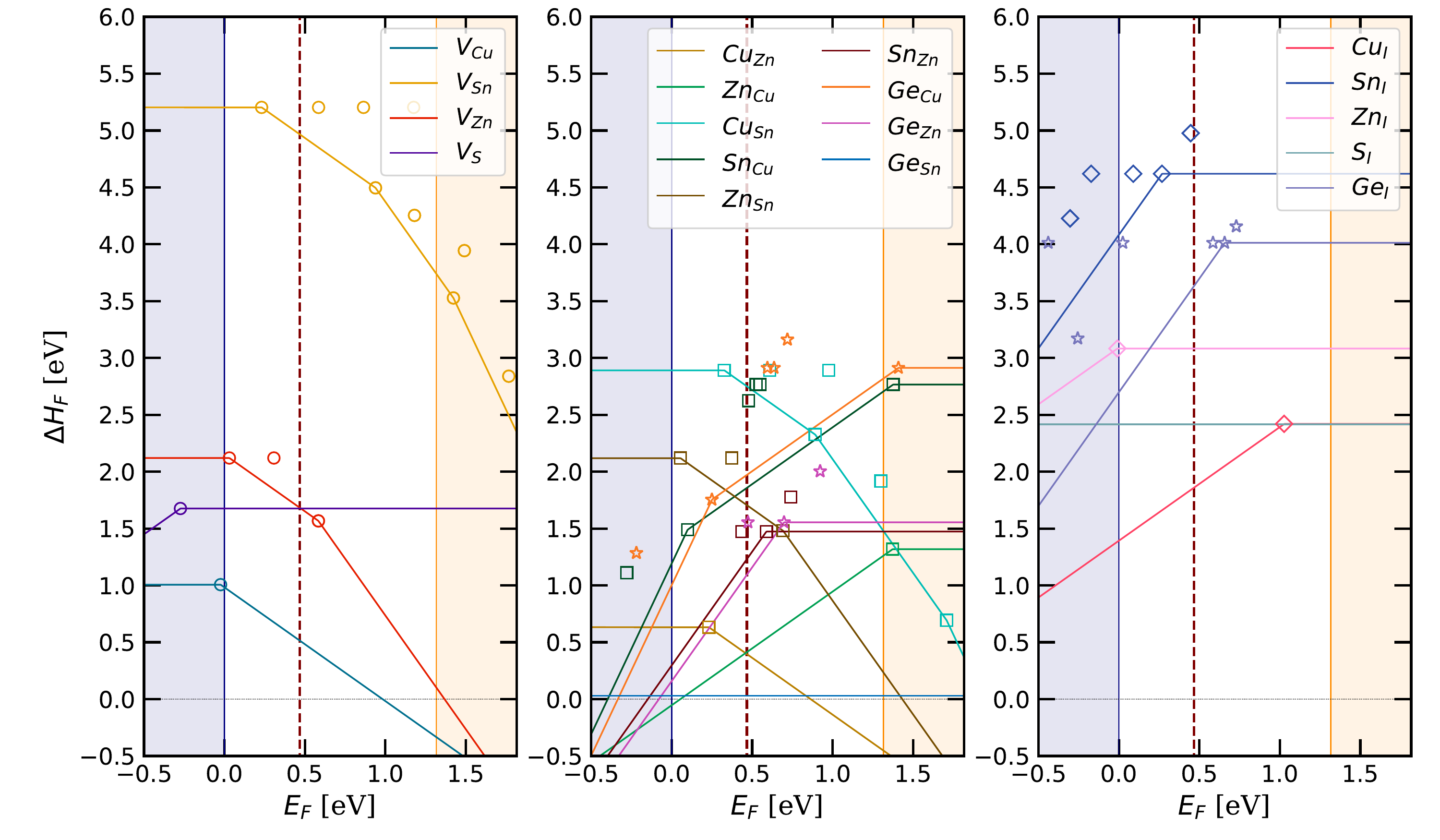}\label{Formation_energy_fig_CZTS}\label{Formation_energy_fig_CZTS}} \\
\subfloat[Cu$_2$ZnGeS$_4$]{\includegraphics[width=0.55\textwidth]{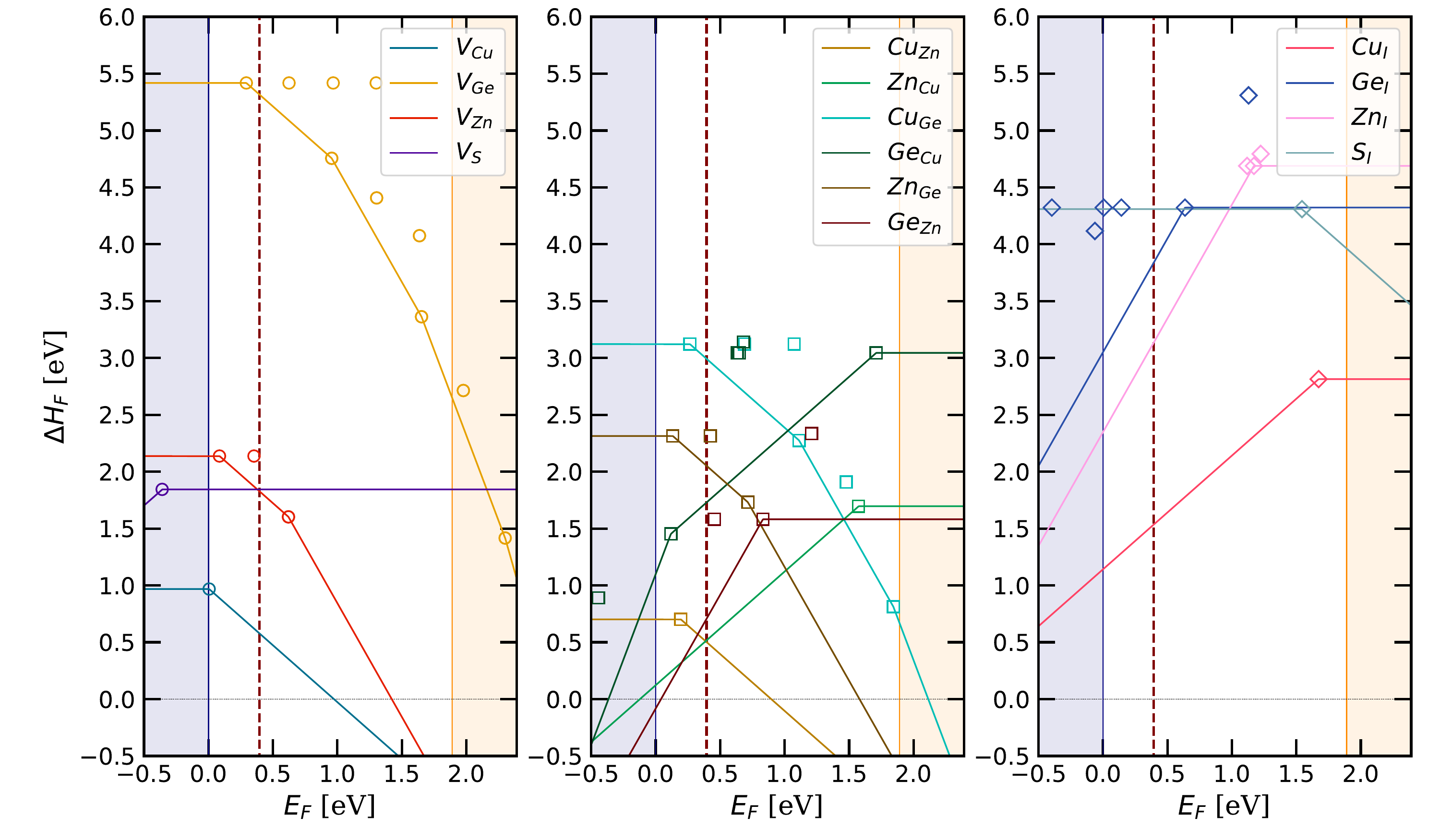}\label{Formation_energy_fig_CZGS}}
\caption{Formation energies of point defects in (a) \czts \ and (b) \czgs \ for three different kinds of defects using a particular marker: vacancy $\mathrm{V_{X}}$ (circle), substitution $\mathrm{X_Y}$ (square) and interstitial $\mathrm{X_I}$ (diamond) calculated at the chemical potential point E corresponding to $\mu_{Cu} = -0.55$ eV, $\mu_{Zn} = -1.56$ eV, $\mu_{Sn} = -0.56$ eV and $\mu_{Ge} = -0.53$ eV for the Sn-kesterite and $\mu_{Cu} = -0.55$ eV, $\mu_{Zn} = -1.51$ eV and $\mu_{Ge} = -0.397$ eV for the Ge-compound. The Fermi level under thermodynamic equilibrium conditions is represented by the maroon dashed lines (see SI). Extrinsic defects corresponding to Ge doping in the Sn-based kesterite matrix are represented using a star marker. A specific color is attributed to each defect and each marker corresponds to a possible transition level between two different charge states of a same defect.}
\label{Formation_energy_fig}
\end{figure*}

In \textbf{Figure \ref{Formation_energy_fig}}, we present the formation energies of kesterite point defects for the chemical potentials $\mu_i$ corresponding to point E in Figure \ref{PD_CZTS_2} and \ref{PD_CZGS_2} (see Equation \ref{Formation_energy_eq}). The choice for this particular composition point was motivated by the perspective of selecting a chemical potential combination corresponding to a pure kesterite phase and as close as possible to the Cu-poor and Zn-rich conditions usually used to synthetise kesterite thin films \cite{Chen:2013cna,Wang:2013gs,schorr2019point}. One has to keep in mind that Figure \ref{Formation_energy_fig} corresponds to a glimpse of the defect formation energy function for one material growth condition. According to Equation \ref{Formation_energy_eq}, the defect formation energy $\Delta H_F (E_F)$ is represented as a function of the Fermi energy level within the kesterite band gaps as predicted by the HOMO-LUMO Kohn-Sham eigenvalues extracted from the first-principles calculations which are equal to 1.32 eV and 1.89 eV respectively for \czts \ and \czgs \ \cite{ratz2021opto}.

Let us first focus on the intrinsic point defect formation energies in the Sn-compound as it is used as our reference material. As shown in Figure \ref{Formation_energy_fig_CZTS}, independently of the Fermi energy value, a first general trend is the lower formation energies of the vacancies (except $\mathrm{V_{Zn}}$) and the substitutional defects in comparison to the interstitial ones. This observation can be explained by the lattice distortion cost induced by the interstitial incorporation. As it can be observed, the lowest formations energies (below 1.5 eV) are reported for the copper vacancy $\mathrm{V_{Cu}}$, and for $\mathrm{Cu_{Zn}}$, $\mathrm{Sn_{Zn}}$ and $\mathrm{Zn_{Cu}}$ substitutional defects. In Figure \ref{Formation_energy_fig_CZGS} we report the same trend concerning the behaviour of the intrinsic point defect formation energies in the Ge-based compound, highlighting the similarity of intrinsic defects in both kesterites.

In addition, as presented in Figure \ref{Formation_energy_fig} and described in the SI, the Fermi level under thermodynamic equilibrium conditions can be extracted. As shown, in both kesterites, under equilibrium conditions the Fermi energy is pinned mainly by the charged defects $\mathrm{V_{Cu}}$ and $\mathrm{Cu_{Zn}}$, both being in a charge state of -1 thus providing holes to the electrical conductivity while the substitutional defect $\mathrm{Zn_{Cu}}$ is in a charge state of +1, supplying electrons. Under equilibrium, the extracted Fermi energy level value is located 0.468 eV and 0.409 eV above the VBM, respectively for \czts \ and \czgs . Through this observation, we consequently highlight the p-type conductivity of \czts , a well-established experimental fact, also confirmed theoretically by Chen \etal \ \cite{Chen:2013cna}. More interestingly, we highlight the same behaviour for the Ge-based kesterite. To complete these observations, we also report the substitutional defect $\mathrm{Sn_{Zn}}$ (resp. $\mathrm{Ge_{Zn}}$) which would be in a charge state +2. However as it will be described later on, $\mathrm{Sn_{Zn}}$ (resp. $\mathrm{Ge_{Zn}}$) would more likely behave as a recombination centre.

\begin{figure}[H]
\centering
\subfloat[Cu$_2$ZnSnS$_4$]{\includegraphics[width=0.5\textwidth]{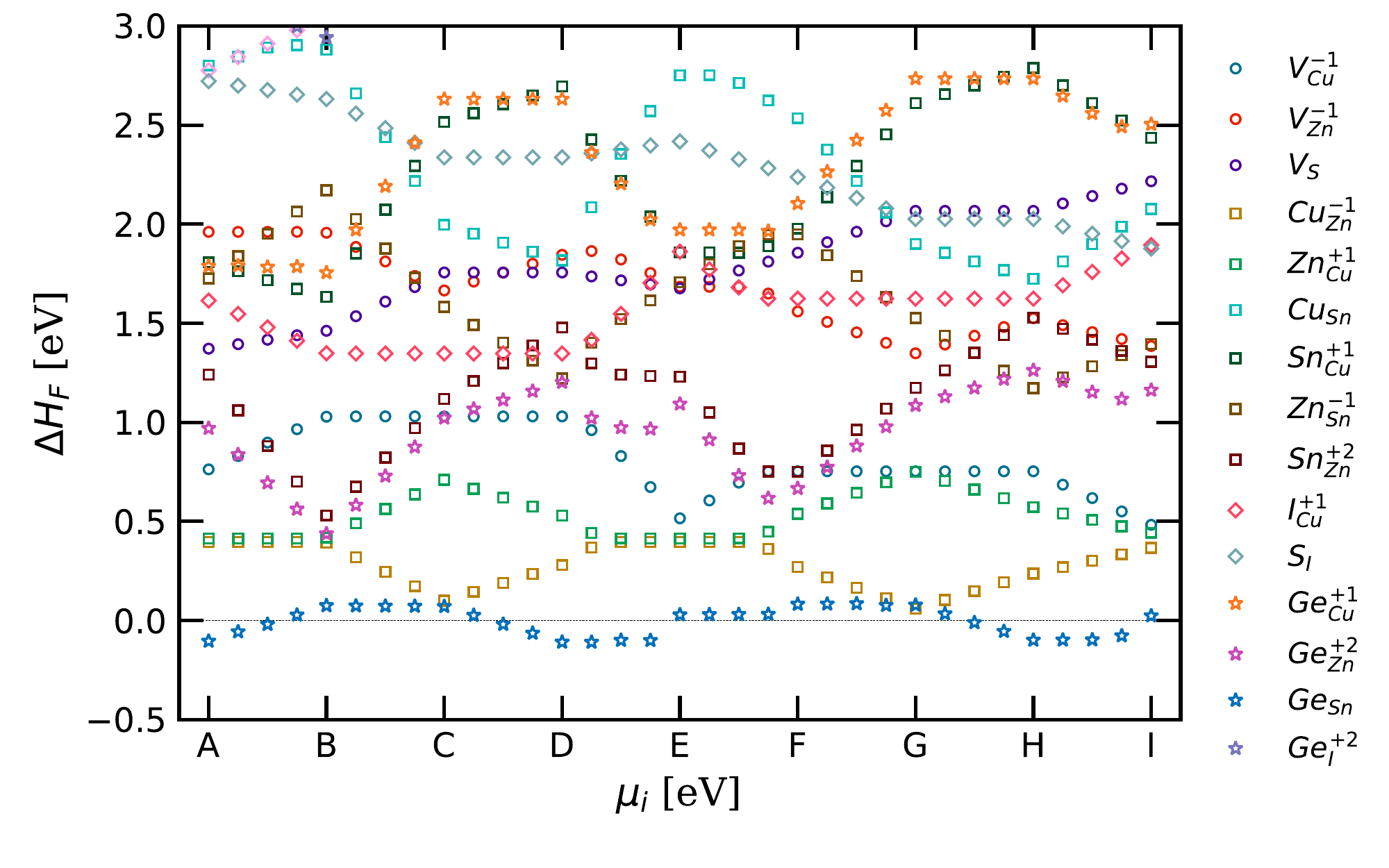}\label{EF_Paths_CZTS}} \hfill
\subfloat[Cu$_2$ZnGeS$_4$]{\includegraphics[width=0.5\textwidth]{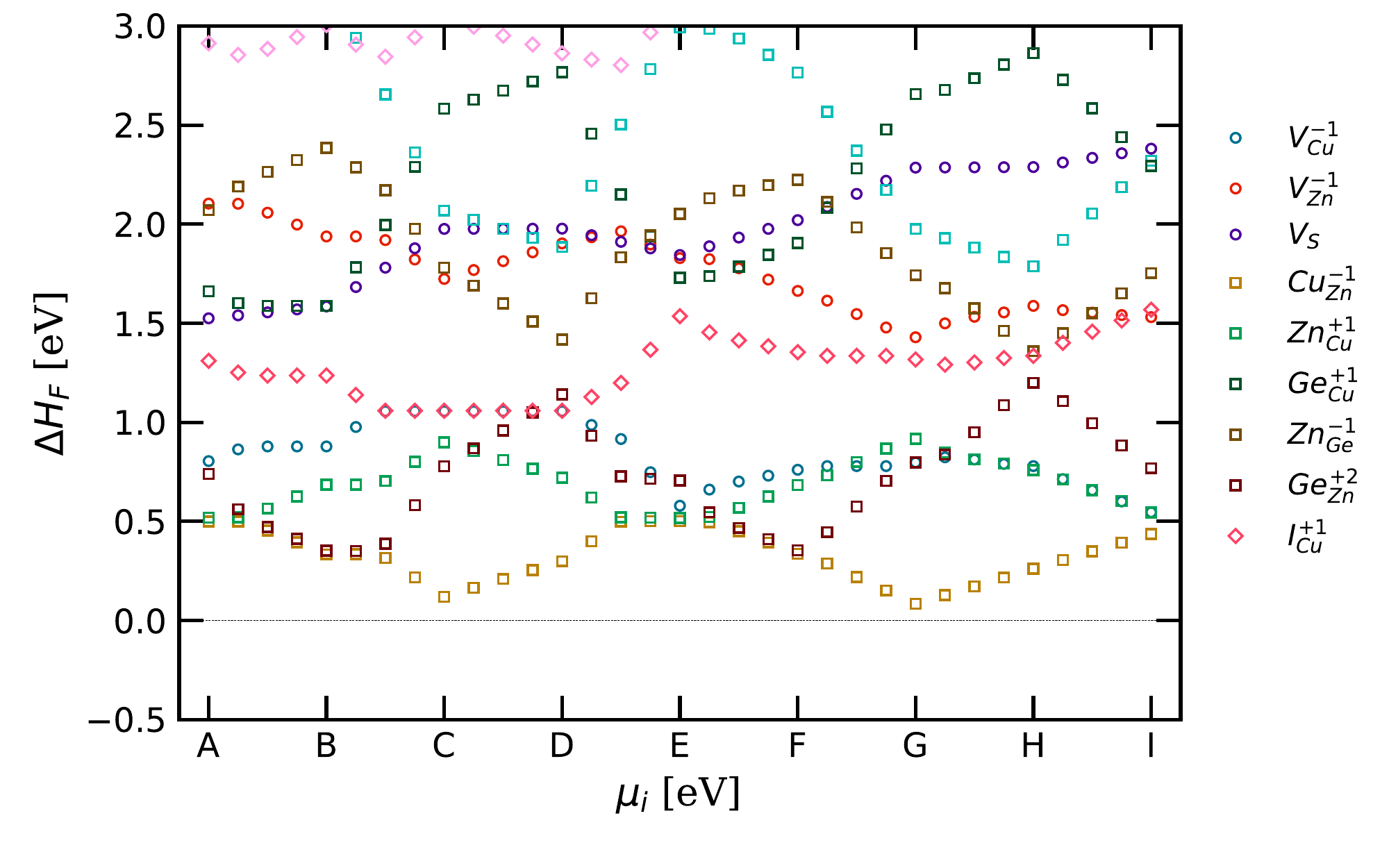}\label{EF_Paths_CZGS}}
\caption{Evolution of the point defect formation energies in (a) \czts \ and (b) \czgs \ for different hypothetical growth conditions. The calculations were performed at various chemical potential combinations corresponding to the label points presented in Figure \ref{PD} (see SI for specific values) and for the Fermi energy levels under thermodynamic equilibrium conditions located at 0.468 eV and 0.409 eV above the VBM respectively for \czts \ and \czgs . Each defect is represented using a specific color and using the following markers: circles for intrinsic vacancies, squares for intrinsic substitutional defects, diamonds for intrinsic interstitials and stars for Ge extrinsic doping defects in the Sn-based kesterite.}
\label{EF_Paths}
\end{figure}

Then, we present Ge-doping extrinsic point defects in the Sn-kesterite matrix such as $\mathrm{Ge_{Cu}}$, $\mathrm{Ge_{Zn}}$, $\mathrm{Ge_{Sn}}$ substitutional defects and $\mathrm{Ge_I}$ interstitial defect. As shown in Figure \ref{Formation_energy_fig_CZTS}, $\mathrm{Ge_{Cu}}$ and $\mathrm{Ge_I}$ have both high formation energies with values above 1.5 eV for any Fermi level value. This result highlights the low probability for the Ge element to be incorporated in the Sn-kesterite via Cu substitution or as interstitial. In contrast, the substitutional defect $\mathrm{Ge_{Zn}}$ presents a formation energy below 1.5 eV, whereas $\mathrm{Ge_{Sn}}$ has a formation energy of nearly zero, meaning that this defect could form spontaneously in presence of Ge. It is also interesting to note that in Figure \ref{EF_Paths_CZTS}, if one compares $\mathrm{Ge_{Cu}}$ and $\mathrm{Sn_{Cu}}$ (resp. $\mathrm{Ge_{Zn}}$ and $\mathrm{Sn_{Zn}}$), both extrinsic doping substitutional defects present similar formation energies to the intrinsic ones. It is also worth noticing than the Ge chemical potential $\mu_{Ge}$ used to compute the Ge-doping defects corresponds to the richest composition value leading consequently to the lowest formation energies. As a result, for a Ge chemical potential value lower than $\mu_{Ge} = -0.53$ eV (poorer composition), the Ge-related defect formation energies will increase (see SI).

However, depending on the growth conditions of the materials, the defect formation energies will vary. Indeed, for a lower concentration of copper during the materials growth, Cu vacancies will form with a greater ease and will consequently be present in a higher concentration. This behaviour is captured by the lowering of the formation energy following the change of the Cu chemical potential value in Equation \ref{Formation_energy_eq}. To visualise these trends, as shown in \textbf{Figure \ref{EF_Paths}}, the defect formation energies were represented for different hypothetical growth conditions. To do so, we represent the evolution of the formation energies following a specific path in the phase diagrams presented in Figure \ref{PD} and for a Fermi energy value under thermodynamic equilibrium conditions (see supplementary information). As presented in the phase diagrams of the Sn and the Ge-kesterite, the chemical potential path study here is labelled as A-B-C-D-E-F-G-H-I. The chemical potentials coordinates A-B-C-D correspond to Cu-rich conditions, E-F-G-H to Cu-moderate conditions and finally the I label corresponds to a Cu-poor condition. This path was selected to study the behaviour of the formation energies at the edges of the kesterite pure phases (as represented via the green shading in Figure \ref{PD}) and following a Cu concentration from high values to lower ones. The relevant message here is consequently focused on the trends of the defect formation energies for different growth conditions and not specifically on the selected path.

In Figure \ref{EF_Paths}, we identify $\mathrm{Cu_{Zn}}^{-1}$, $\mathrm{Zn_{Cu}}^{+1}$, $\mathrm{Sn_{Zn}}^{+2}$ (resp. $\mathrm{Ge_{Zn}}^{+2}$) and $\mathrm{V_{Cu}}^{-1}$ as the intrinsic point defects showing the lowest formation energies over the different chemical potential combinations with values below 1.5 eV. Following these results, the usually observed Cu/Zn disorder in Sn-kesterite can also be expected in the Ge-kesterite as the two substitutional defects $\mathrm{Cu_{Zn}}^{-1}$ and $\mathrm{Zn_{Cu}}^{+1}$ present formation energies below 1 eV for any chemical potential combination. This result is in good agreement with those of Chen \etal \ where $\mathrm{Cu_{Zn}}^{-1}$ is the dominant point defect \cite{Chen:2013cna}. In addition, we report a lower formation energy concerning the $\mathrm{Zn_{Cu}}^{+1}$ antisite with a formation energy between 0.03 eV and 0.63 eV while this previous work predicted a formation energy between 0.6 and 0.9 eV. As a results, our calculations predict a subtitutional defect $\mathrm{Zn_{Cu}}^{+1}$ concentration within the same range as the $\mathrm{Cu_{Zn}}^{-1}$ defect. This observation is in good agreement with the work of Du \etal \ reporting a similar $\mathrm{Zn_{Cu}}^{+1}$ behaviour \cite{du2021defect}. The facilitated formation of these defects could be explained by the similar atomic radii of Zn ($r_{\mathrm{Zn}}$ = 1.35 $\angstrom$) and Cu ($r_{\mathrm{Zn}}$ = 1.35 $\angstrom$) while their electronic configurations differ only by one electron \cite{slater1964atomic}. Several studies hold the related Cu/Zn disorder defects accountable for potential fluctuations detrimental for solar cell performances \cite{ma2019origin, rey2018origin} which could however be suppressed via Ag incorporation \cite{he2021systematic}.

As shown in Figure \ref{EF_Paths_CZTS}, focusing on the Ge extrinsic doping defects, the substitutional defect $\mathrm{Ge_{Sn}}$ presents the lowest formation energies of all for every chemical potential combinations. The smaller atomic radius of Ge ($r_{\mathrm{Ge}} = 1.25 \angstrom$) in comparison to the Sn element ($r_{\mathrm{Sn}} = 1.45 \angstrom$) and their similar electronic behaviours could explain the high occurrence of this defect. In contrast, $\mathrm{Ge_{Zn}^{+2}}$ presents a higher formation energy which is equal or a few tenths of eV lower than the intrinsic $\mathrm{Sn_{Zn}}$ substitutional defect. We finally report the case of $\mathrm{Ge_{Cu}^{+1}}$ and $\mathrm{Ge_{I}^{+2}}$ with formation energies above 1.5 eV. Moving from the Ge doping to the Ge alloying (see Figure \ref{EF_Paths_CZGS}), for a fixed chemical potential combination (\textit{i.e.} E point), a decrease of the formation energy is observed from the intrinsic defect $\mathrm{Sn_{Zn}}^{+2}$ ($\Delta H_F = 1.23$ eV) in the Sn-kesterite to the $\mathrm{Ge_{Zn}}^{+2}$ ($\Delta H_F = 1.09$ eV) extrinsic doping defect in the Sn-compound ultimately to the intrinsic $\mathrm{Ge_{Zn}}^{+2}$ ($\Delta H_F = 0.706$ eV) defect in the Ge material. As a result, it would appear that if Ge is available, the facilitated formation of the Zn substitutional defect would be further improved, first in the Sn-kesterite by doping and then in the Ge-kesterite by alloying. 

Here, we have reported $\mathrm{Cu_{Zn}}^{-1}$, $\mathrm{Zn_{Cu}}^{+1}$, $\mathrm{Sn_{Zn}}^{+2}$ (resp. $\mathrm{Ge_{Zn}}^{+2}$) and $\mathrm{V_{Cu}}^{-1}$ as the most abundant point defects in \czts \ (resp. \czgs ). In the next section, we focus on the physical behaviour of the following ones: acceptors, donors or recombination centres.

\subsection{Defect identification: ionisation levels}

\begin{figure}[H]
\centering
\subfloat[Cu$_2$ZnSnS$_4$]{\includegraphics[width=0.53\textwidth]{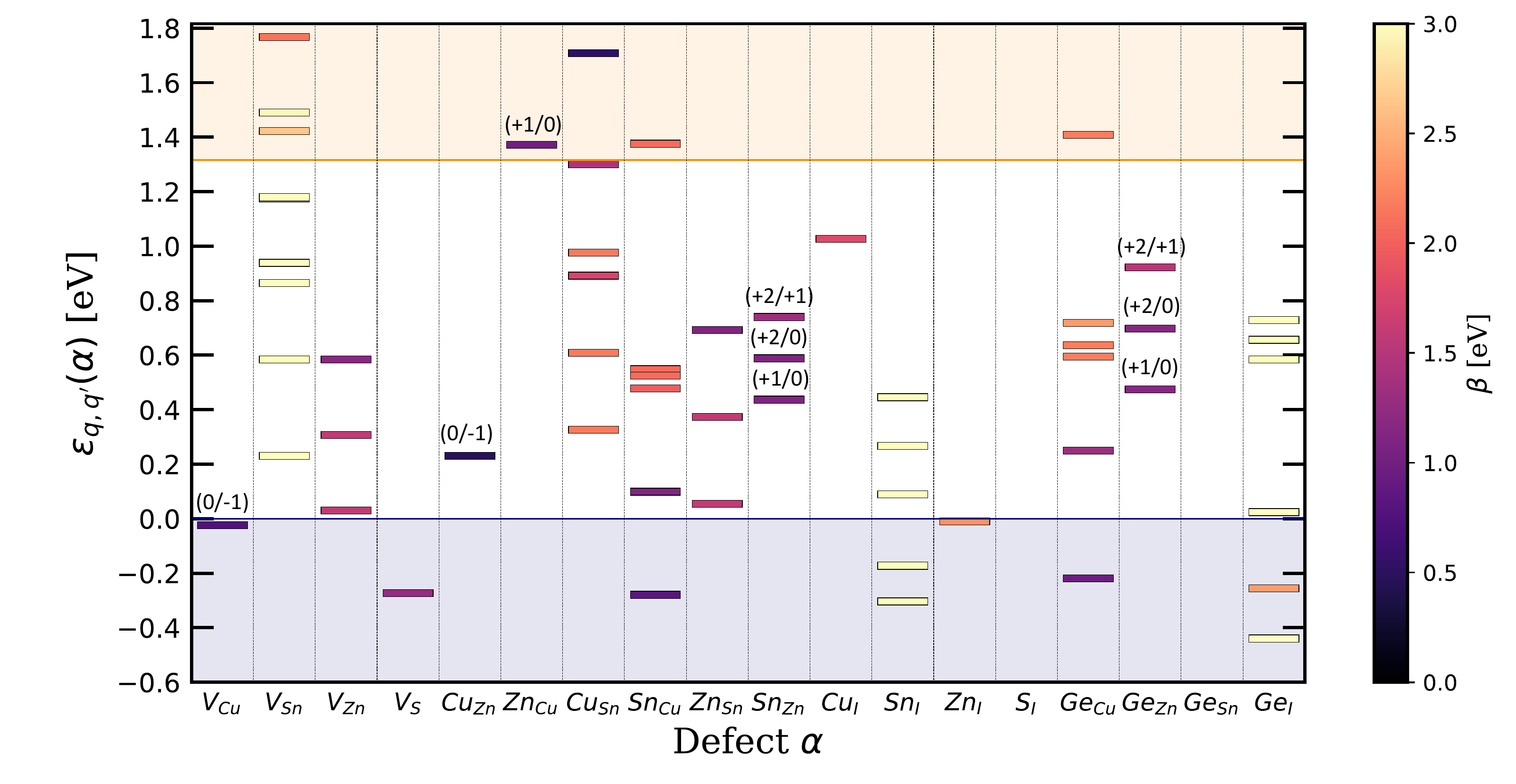}\label{Atom_ionisation_levels_CZTS}} \hfill
\subfloat[Cu$_2$ZnGeS$_4$]{\includegraphics[width=0.455\textwidth]{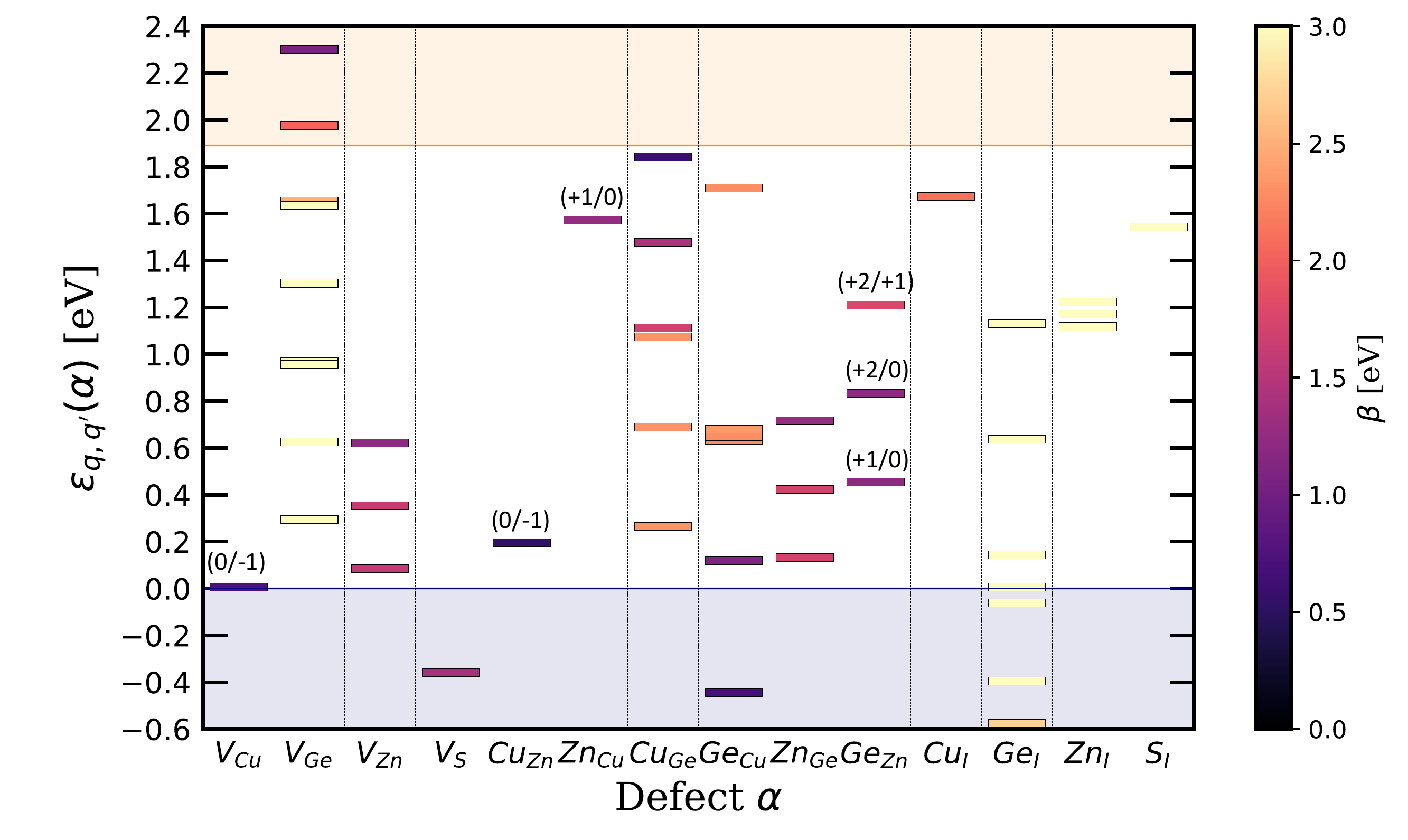}\label{Atom_ionisation_levels_CZGS}}
\caption{Ionisation energy levels of intrinsic point defects in (a) \czts \ and (b) \czgs \ calculated for a chemical potential combination corresponding to point E in Figure \ref{PD_CZTS_2} and Figure \ref{PD_CZGS_2}. Their locations within the kesterite band gaps are reported here on the y-axis while the $\beta$ value corresponds to the formation energy value at which the transition occurs in Figure \ref{Formation_energy_fig}. This $\beta$ value is provided as a guide for the eye to highlight the most dominant transition levels.}
\label{Atom_ionisation_levels}
\end{figure}

In \textbf{Figure \ref{Atom_ionisation_levels}}, we show the ionisation levels for each defect according to Equation \ref{ionization_eq}. It is important to note that these levels within the kesterite band gap are, in contrast to the formation energies, independent of the chemical potential values. The relevant information here are (i) the position of the ionisation level in the material band gap, which determines the behaviour of the defect, and (ii) as a guide for the eye, the "formation energy value" of the ionisation energy level $\beta$ for a Fermi level located at the transition level within the kesterite band gap (see also markers in Figure \ref{Formation_energy_fig}). 

For both \czts \ and \czgs , as shown Figure \ref{Atom_ionisation_levels}, we highlight the (0/-1) transition levels of $\mathrm{Cu_{Zn}}$ and $\mathrm{V_{Cu}}$ located a few $k_B T$ over the VBM. From the calculation of the Fermi energy under themodynamic equilibrium condition, these transition levels are ionised and consequently, these defects act as acceptors both providing holes to the kesterite conductivity. Combined with their low formation energies, these levels are particularly suited to account for the p-type intrinsic conductivity of these compounds. Then, as donor defects, we report the (+1/0) transition level of the substitutional defect $\mathrm{Zn_{Cu}}$ located close to the conduction band minimum (CBM). With its previously reported low formation energy, this defect is the most abundant donor in both kesterites. We can also note that in the case of the Ge-compound, in opposition to the Sn-kesterite, the (+1/0) transition level of $\mathrm{Zn_{Cu}}$ is below the CBM. This is a direct consequence of the band gap increase associated to the Ge alloying. The pinning of the Fermi level under thermodynamic equilibrium condition is a result of these two observations, which explain the p-type conductivity of kesterite materials generally reported in the literature \cite{Chen:2013cna}. We also notice that the $\mathrm{V_{Zn}}$ defect provides energy levels close to the VBM, however, assuming a p-type material, the Fermi level would be close to the valence band leading to an unionised defect with a high formation energy as presented in Figure \ref{Formation_energy_fig_CZTS}. Concerning the substitutional defect $\mathrm{Sn_{Cu}}$ (resp. $\mathrm{Ge_{Cu}}$), they present ionisation states not suited for intrinsic doping (\textit{i.e.} "donor states" close to the valence band). A similar behaviour is observed for the substitutional defect $\mathrm{Cu_{Sn}}$. As a consequence, these defects should not contribute significantly to the conductivity in any of the two kesterites.

One can identify $\mathrm{Sn_{Zn}}$ (resp. $\mathrm{Ge_{Zn}}$), $\mathrm{Zn_{Sn}}$ (resp. $\mathrm{Zn_{Ge}}$) and to a lesser extent $\mathrm{Cu_{Sn}}$ (resp. $\mathrm{Cu_{Ge}}$) and $\mathrm{Sn_{Cu}}$ (resp. $\mathrm{Ge_{Cu}}$) as deep defects. Indeed, all these substitutional defects offer transition levels located close to the middle of the kesterite band gap. In addition, as presented in Figure \ref{EF_Paths}, the substitution of Zn by Sn (resp. Ge) is the only defect presenting also low formation energy (below 1.5 eV). The latter has several transition levels within the kesterite band gap corresponding to the various transitions between its various charge states: (+2/0) at 0.584 eV (resp. 0.736 eV), (+2/+1) at 0.735 eV (resp. 0.585 eV) and (+1/0) at 0.43 eV (resp. 0.890 eV) above the VBM (below the CBM). These values are comparable to those reported by Li \etal \ with $\mathrm{Sn_{Zn}}$ (+1/0) at 0.86 eV and (+2/0) at 0.67 eV  below the CBM \cite{li2019effective}. Moreover, as a result of the p-type conductivity, the $\mathrm{Sn_{Zn}}$ (resp. $\mathrm{Ge_{Zn}}$) defect should be ionised into a charge state +2, with the most probable transition energy level being the two-electron transition (+2/0)  (see Figure \ref{Formation_energy_fig}). Using deep-level transient spectroscopy, Deng \etal \ reported a defect activation energy of 0.581 eV for the Sn-based kesterite, and identified this defect as a recombination centre \cite{deng2021adjusting}. As shown in Figure \ref{Atom_ionisation_levels_CZTS} and in agreement with this experimental observation, this defect could be reported as the substitutional defect $\mathrm{Sn_{Zn}}$ with its two-electron transition level (+2/0) at 0.584 eV above the VBM. Furthermore, the same authors reported a decrease of the transition level position to 0.542 eV following Ge incorporation, while in this work, upon Ge incorporation we report the spreading of the ionisation levels located between 0.42 eV and 0.73 eV for $\mathrm{Sn_{Zn}}$, between 0.46 eV and 0.91 eV for $\mathrm{Ge_{Zn}}$ (in Sn-based kesterite) and between 0.44 eV and 1.9 eV for $\mathrm{Ge_{Zn}}$ (in Ge-based kesterite). It should also be noted that in all three situations, the transition level (+2/0) is the closest to the middle of the gap.

Beyond, formation energies, other physical parameter can also act as indicators for the evaluation of the impact of point defects on solar cell properties. In the following section, we consider recombination centre capture cross sections, which can be related to the kesterite lattice distortion upon introduction of the defect.

\subsection{Atomic distortions}

\begin{figure}[H]
\centering
\includegraphics[width=0.5\textwidth]{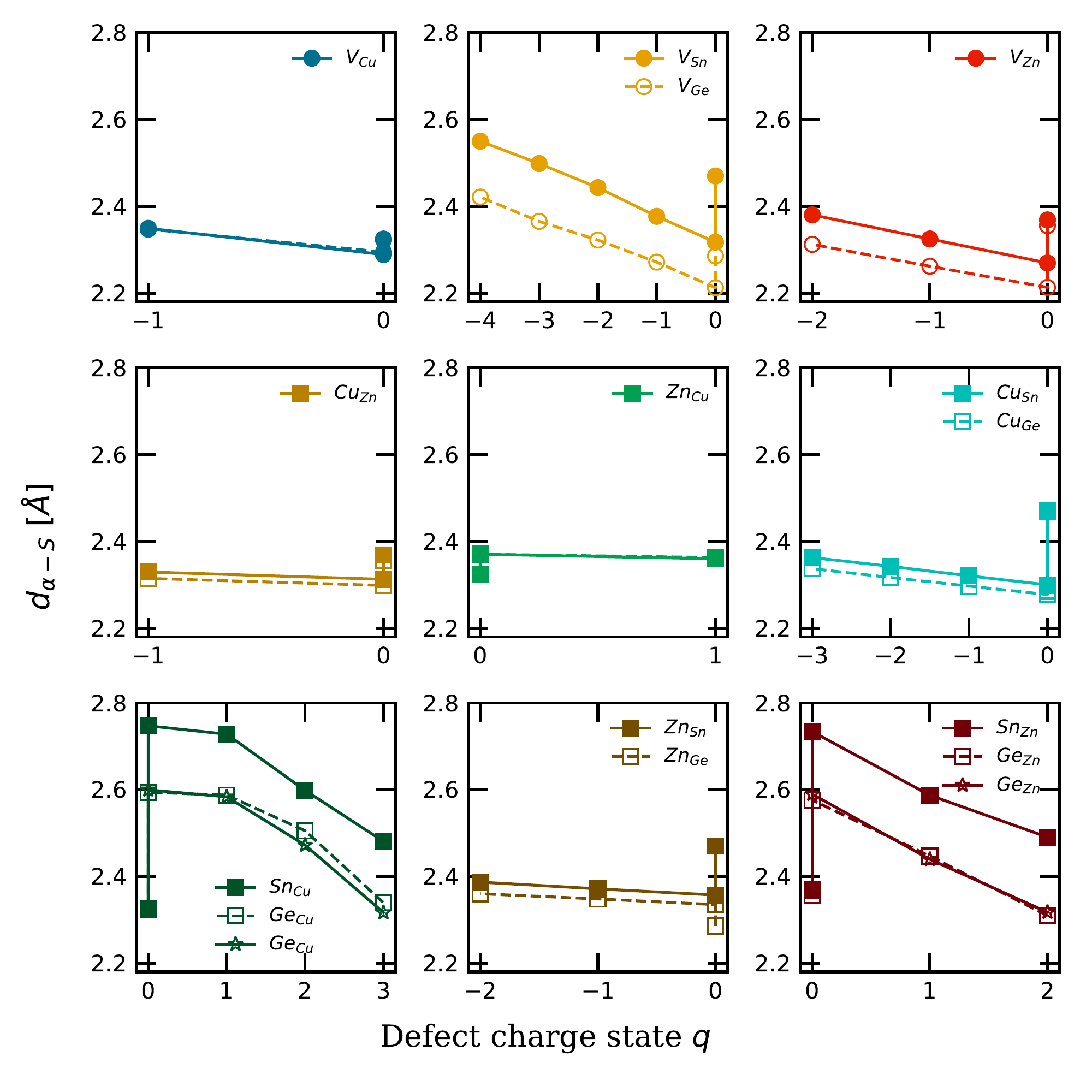} \\
\caption{Representation of the interatomic distance between the defect position and the surrounding S atoms with respect to the defect charge state $q$. For a charge state $q=0$, the evolution corresponds to the distortion undergone by the lattice following the defect incorporation while the evolution along the x-axis corresponds to the lattice distortion as a result of the capture/emission of electrons by the defect. The results corresponding to the \czts \ (resp. \czgs ) compound are reported in plain (resp. dashed) lines. Ge-doping distortions in Sn-kesterite are reported in plain lines using a star symbol.}
\label{Distortion_fig}
\end{figure}

In \textbf{Figure \ref{Distortion_fig}}, we present the evolution of the interatomic distances between the defect position and the surrounding S atoms with respect to the charge states $q$ of the defect, both in the Sn-based and the Ge-based compounds. As reported by Li \etal \ \cite{li2019effective}, an empirical approach to evaluate the carrier capture cross section of defects consists in studying the local structural relaxation undergone by the lattice when the defect captures/emits electron(s) \cite{li2019effective}. They suggested that strong bonds and large structural relaxations imply large defect capture cross sections. Such a qualitative approach can be more firmly grounded using a quantitative study of the defect carrier capture cross section via the computation of the phonon-electron Hamiltonian \cite{alkauskas2014first, li2019effective}. As expressed in Equation \ref{capturerate}, the emission/capture rate of a defect does not depend solely on the capture cross section. Indeed, the electron emission rate $e_n$ (resp. $e_p$ for holes) is related to its ionisation level in the kesterite band gap with respect to the CBM: $E_t - E_C $ (resp. to the VBM: $E_t-E_V$ for holes) and to the capture cross section of the defect $\sigma_n$ (resp. $\sigma_p$) following this relation:

\begin{equation}
e_n = \sigma_n \langle v_n \rangle N_C \ \mathrm{exp} \bigg( - \frac{E_C-E_t}{k_B T} \bigg) , 
\label{capturerate}
\end{equation}
where $\langle v_n \rangle$ (resp. $\langle v_p \rangle$) is the average thermal velocity of the electron (resp. hole) and $N_C$ (resp. $N_V$) the effective density of states in the conduction band (resp. valence band).

A first general trend that can be observed for each defect is that as the charge state of the defect gets closer to the electronic configuration of the pristine system, the lattice distortion tends to get closer to the reference distance. For example, in the case of a substitutional defect, if both elements have similar atomic radii and if their electronic configurations are close, one could expect a small distortion undergone bye the lattice upon defect incorporation.
 
A similar behaviour is observed in both \czts \ and \czgs \ concerning defects providing transition levels close to the band gap edges and showing low formation energies ($\mathrm{Cu_{Zn}}$, $\mathrm{Zn_{Cu}}$ and $\mathrm{V_{Cu}}$). The introduction of the copper vacancy leads only to a small distortion of the lattice as the interatomic distance variation is less than 0.1 \AA . In addition, assuming the capture of an electron by such a defect, the value of the interatomic distance remains quite constant in comparison to the one in the pristine lattice. This result supports the readiness of these defects to provide charge carriers in kesterite materials. We report the same behaviour for the substitutional defects $\mathrm{Cu_{Zn}}$ and $\mathrm{Zn_{Cu}}$. The results also support that the Cu/Zn disorder commonly observed in Sn-based kesterite should be present as well in the Ge-based compound. In addition, for the $\mathrm{Ge_{Sn}}$ substitutional defect in the Ge-based kesterite, we report a reduction of the X-S (X=Sn,Ge) interatomic distance from 2.46 \AA \ in the Sn-based compound to 2.30 \AA \ in the Ge-based material. This can be interpreted as a direct consequence of the smaller atomic raduis of Ge ($r_{Ge} = 1.25$ \AA \ compare to the Sn element: $r_{Sn} = 1.45$ \AA \ \cite{slater1964atomic}).

Moreover, we observe the largest lattice distortions for both substitutional defects $\mathrm{Sn_{Cu}}$ (resp. $\mathrm{Ge_{Cu}}$) and $\mathrm{Sn_{Zn}}$ (resp. $\mathrm{Ge_{Zn}}$) which were identified as possible recombination centres in the previous section. In good agreement with our calculations, Li \etal \ reported similar values concerning the Sn-S distances for various charged states of the $\mathrm{Sn_{Zn}}$ defect, \textit{i.e.} 2.71 \AA \ ($q=0$), 2.57 \AA \ ($q=1$) and 2.43 \AA \ ($q=2$) while the values obtained in this work are 2.73 \AA , 2.59 \AA \ and 2.49 \AA \, respectively for $q=0, 1 $ and 2. In addition, the distortion inside the Sn-kesterite is stronger than that found in the Ge-based kesterite. Following the empirical rule proposed by Li \etal \ in Ref.\cite{li2019effective}, this would lead to a smaller carrier capture cross section in the Ge-compound, partially explaining the smaller $V_{OC}$ deficit reported in the Ge-based kesterite with respect to the Sn-based material.

To summarise, following the Ge incorporation in the Sn-kesterite, from Ge-doping to pure Ge-alloying material, we report a decrease of the $\mathrm{X_{Zn}}$ (X=Sn,Ge) substitutional defect formation energies for any chemical potential combination which consequently results in an increase of the defect concentration. Secondly, along with this observation, the ionisation levels associtated to the $\mathrm{X_{Zn}}$ (X=Sn,Ge) defect tend to spread within the material band gap as we move from Ge doping to Ge alloying with the (+2/0) transition level still located in the middle of the gap. Finally, from the lattice distortion investigation, we report a reduction of the $\mathrm{X_{Zn}-S}$ (X=Sn,Ge) interatomic distance, which results in a decrease of the associated carrier capture cross section. In addition to these observations, we previously reported an increase of the $V_{OC}$ value as well as a decrease of the $J_{SC}$ value as a result of the increase of the material band gap from 1.32 eV (\czts) to 1.89 eV (\czgs) \cite{ratz2021opto}. Combining the results from both investigations, we thus conclude that the improvement of the solar cell efficiency associated to Ge doping of Sn-based kesterites is ascribed to the improvement of the $V_{OC}$ value while maintaining the $J_{SC}$ value of Sn-based kesterite cell. This enhancement is the result of the smaller carrier capture cross section of $\mathrm{Ge_{Zn}}$ defects in comparison to their Sn counterpart. This reduction appears to be dominant with respect to the reported increasing of the defect concentration.

\section{Conclusion}

This work was devoted to first-principles investigations of Ge-related defects in kesterite. First, we highlight the slightly wider pure phase range of chemical potential values of the Ge-based kesterite compared to the Sn-based kesterite. In both cases, the pure phase remains limited in terms of possible chemical potential ranges due to the numerous secondary phases, which are a direct consequence of all four elements present in these kesterites. Near the stable phase region, the secondary phases are ZnS, CuS and Cu$_2$SnS$_3$ (resp. Cu$_2$GeS$_3$) in the Sn-kesterite (resp. Ge-kesterite). 
We found a similar physical behaviour of the intrinsic point defects for the Sn-kesterite and the Ge-kesterite. In both compounds, we identified $\mathrm{V_{Cu}^{-1}}$, $\mathrm{Cu_{Zn}^{-1}}$ and $\mathrm{Zn_{Cu}^{+1}}$ as low formation energy defects acting as acceptors and as donor defect for the latter. By calculating the Fermi level under equilibrium conditions, we confirmed the p-type conductivity reported in the literature for both the Sn-based and the Ge-based compounds. We also shed a light on the commonly observed Cu/Zn disorder encountered in kesterite compounds. 
In addition, via the study of Ge doping in the Sn-kesterite compound, we identified the $\mathrm{Ge_{Sn}}$ neutral defect as a spontaneous defect, an indication that the Ge doping within the kesterite matrix occurs via a Sn substitution. 
Finally, we identified $\mathrm{Sn_{Zn}}$ and $\mathrm{Ge_{Zn}}$ as recombination centres. We reported a decrease of the substitutional defect formation energy following the Ge-doping and alloying which would result in an increase of the defect concentration. Moreover, it appeared that the lattice distortion induced by the formation of these defects is reduced for the $\mathrm{Ge_{Zn}}$ substitutional defect in comparison to its Sn counterpart. This result hints at a reduced carrier capture cross section and consequently a less detrimental defect behaviour to be ascribed to $\mathrm{Ge_{Zn}}$ with respect to $\mathrm{Sn_{Zn}}$. As a consequence, we pointed out the reduction of the non-radiative recombination rate induced by the reduction of the carrier capture cross section of the detrimental defect $\mathrm{X_{Zn}}$ (X=Sn,Ge) as one of the sources of the $V_{OC}$ improvement reported in the literature upon Ge incorporation.

The objective of this work was to strengthen the understanding of the effects of Ge doping and Ge alloying in kesterite materials for photovoltaic applications. We believe that our results clarified the fundamental mechanisms that operate at the atomic scale via the formation of a wide range of point defects and linked them to photovoltaic properties enhancement of Ge-based kesterite reported in previous works. 

% Experimental section

\section{Method}
\threesubsection{Computational method}\\

To compute the total energies of the defected and host supercells (as required in Equation\ref{Formation_energy_eq}) and of the secondary phases and pure elemental phases (see supplementary materials), first-principles calculations have been performed using the Vienna \textit{Ab initio} Simulation Package (VASP) code \cite{kresse1996efficiency} with the Projector-Augmented Wave (PAW) potential method \cite{kresse1999ultrasoft}. Perdew-Burke-Ernzerhof (PBE) GGA pseudo-potentials \cite{perdew1996generalized} were used with the following valence electrons considered for each element: Cu: 3d$^{10}$4s$^1$, Zn: 3d$^{10}$4s$^2$, Sn: 4d$^{10}$5s$^2$5p$^2$, Ge: 3d$^{10}$4s$^2$4p$^2$ and S: 3s$^2$3p$^4$. Based on the previously presented supercell approach, within a 64-atoms supercell, ionic and electronic relaxations were achieved using a cut-off energy of 520 eV and a $\Gamma$-centered uniform \textbf{k}-points mesh of $2 \times 2 \times 2$ \textbf{k}-points. Applying the strongly constrained and appropriately normed semilocal density functional (SCAN) \cite{sun2015strongly, sun2016accurate}, the structures were relaxed until the numerical convergence regarding the self-consistent cycles reached forces between ions less than $10^{-3}$ eV/$\angstrom$. Starting from the relaxed structures, a single-shot calculation using the Heyd–Scuseria–Ernzerhof exchange-correlation functional (HSE06) \cite{heyd2003hybrid} is performed using an energy convergence criterion upon $10^{-3}$ eV. The combination of SCAN ionic relaxations followed by a single HSE06 electronic relaxation was reported as an efficient method to obtain accurate results by Fritsch \etal \ in Ref.\cite{fritsch2020climbing}.

% Acknowledgements
\medskip
\textbf{Acknowledgements} \par %delete if not applicable))
Computational resources have been provided by the Consortium des Équipements de Calcul Intensif (CÉCI), funded by the Fonds de la Recherche Scientifique de Belgique (F.R.S.-FNRS) under Grant No. 2.5020.11 and by the Walloon Region. JY thanks the FNRS for the ABIGLO grant J.0154.21.

% References
\medskip
% Use the following code if you wish to generate your bibliography with BibTeX;
% replace the string "MSP-template" below with the name(s) of
% the BibTeX data base(s) you want to use.
% The resulting bibliography-output (the content of the .bbl file)
% must be pasted back into this file before submission.
% Please also include your BibTeX data base file(s) in your submission
% so that we can re-run BibTeX if necessary.
\bibliographystyle{MSP}
\bibliography{CZTS_defect_bib}

\begin{thebibliography}{10}
\providecommand{\url}[1]{\texttt{#1}}
\providecommand{\urlprefix}{URL }

\bibitem{Green:2020ga}
M.~Green, E.~Dunlop, J.~Hohl-Ebinger, M.~Yoshita, N.~Kopidakis, X.~Hao,
\newblock \emph{Progress in Photovoltaics: Research and Applications}
  \textbf{2021}, \emph{29}, 1 3.

\bibitem{nakamura2019cd}
M.~Nakamura, K.~Yamaguchi, Y.~Kimoto, Y.~Yasaki, T.~Kato, H.~Sugimoto,
\newblock \emph{IEEE Journal of Photovoltaics} \textbf{2019}, \emph{9}, 6 1863.

\bibitem{Giraldo:2019iia}
S.~Giraldo, Z.~Jehl, M.~Placidi, V.~Izquierdo-Roca,
  A.~P{\'e}rez-Rodr{\'\i}guez, E.~Saucedo,
\newblock \emph{Advanced Materials} \textbf{2019}, \emph{31}, 16 1806692.

\bibitem{Ratz:2019cs}
T.~Ratz, G.~Brammertz, R.~Caballero, M.~Le{\'o}n, S.~Canulescu, J.~Schou,
  L.~G{\"u}tay, D.~Pareek, T.~Taskesen, D.-H. Kim, J.~K. Kang, C.~Malerba,
  A.~Redinger, E.~Saucedo, B.~Shin, H.~Tampo, K.~Timmo, N.~D. Nguyen,
  B.~Vermang,
\newblock \emph{Journal of Physics: Energy} \textbf{2019}, \emph{1}, 4 042003.

\bibitem{todorov2020solution}
T.~Todorov, H.~Hillhouse, S.~Aazou, Z.~Sekkat, O.~Vigil-Gal{\'a}n, S.~Deshmukh,
  R.~Agrawal, S.~Bourdais, M.~Vald{\'e}s, P.~Arnou, et~al.,
\newblock \emph{Journal of Physics: Energy} \textbf{2020}, \emph{2}, 1 012003.

\bibitem{Chen:2013cna}
S.~Chen, A.~Walsh, X.-G. Gong, S.-H. Wei,
\newblock \emph{Advanced materials} \textbf{2013}, \emph{25}, 11 1522.

\bibitem{Kim:2018jd}
S.~Kim, J.-S. Park, A.~Walsh,
\newblock \emph{ACS Energy Letters} \textbf{2018}, \emph{3}, 2 496.

\bibitem{grossberg2019electrical}
M.~Grossberg, J.~Krustok, C.~J. Hages, D.~M. Bishop, O.~Gunawan, R.~Scheer,
  S.~M. Lyam, H.~Hempel, S.~Levcenco, T.~Unold,
\newblock \emph{Journal of Physics: Energy} \textbf{2019}, \emph{1}, 4 044002.

\bibitem{pal2019current}
K.~Pal, P.~Singh, A.~Bhaduri, K.~B. Thapa,
\newblock \emph{Solar Energy Materials and Solar Cells} \textbf{2019},
  \emph{196} 138.

\bibitem{dimitrievska2016secondary}
M.~Dimitrievska, A.~Fairbrother, E.~Saucedo, A.~P{\'e}rez-Rodr{\'\i}guez,
  V.~Izquierdo-Roca,
\newblock \emph{Solar Energy Materials and Solar Cells} \textbf{2016},
  \emph{149} 304.

\bibitem{schorr2019point}
S.~Schorr, G.~Gurieva, M.~Guc, M.~Dimitrievska, A.~P{\'e}rez-Rodr{\'\i}guez,
  V.~Izquierdo-Roca, C.~S. Schnohr, J.~Kim, W.~Jo, J.~M. Merino,
\newblock \emph{Journal of Physics: Energy} \textbf{2019}, \emph{2}, 1 012002.

\bibitem{PlatzerBjorkman:2019ed}
C.~Platzer-Bj{\"o}rkman, N.~Barreau, M.~B{\"a}r, L.~Choubrac, L.~Grenet,
  J.~Heo, T.~Kubart, A.~Mittiga, Y.~S{\'a}nchez, J.~Scragg, S.~Sinha,
  M.~Valentini,
\newblock \emph{Journal of Physics: Energy} \textbf{2019}, \emph{1}, 4 044005.

\bibitem{crovetto2017band}
A.~Crovetto, O.~Hansen,
\newblock \emph{Solar Energy Materials and Solar Cells} \textbf{2017},
  \emph{169} 177.

\bibitem{gokmen2013band}
T.~Gokmen, O.~Gunawan, T.~K. Todorov, D.~B. Mitzi,
\newblock \emph{Applied Physics Letters} \textbf{2013}, \emph{103}, 10 103506.

\bibitem{biswas2010electronic}
K.~Biswas, S.~Lany, A.~Zunger,
\newblock \emph{Applied Physics Letters} \textbf{2010}, \emph{96}, 20 201902.

\bibitem{gong2021identifying}
Y.~Gong, Y.~Zhang, Q.~Zhu, Y.~Zhou, R.~Qiu, C.~Niu, W.~Yan, W.~Huang, H.~Xin,
\newblock \emph{Energy \& Environmental Science} \textbf{2021}, \emph{14}, 4
  2369.

\bibitem{Romanyuk:2019cq}
Y.~E. Romanyuk, S.~G. Haass, S.~Giraldo, M.~Placidi, D.~Tiwari, D.~J. Fermin,
  X.~Hao, H.~Xin, T.~Schnabel, M.~Kauk-Kuusik, P.~Pistor, S.~Lie, L.~H. Wong,
\newblock \emph{Journal of Physics: Energy} \textbf{2019}, \emph{1}, 4 044004.

\bibitem{crovetto2020assessing}
A.~Crovetto, S.~Kim, M.~Fischer, N.~Stenger, A.~Walsh, I.~Chorkendorff, P.~C.
  Vesborg,
\newblock \emph{Energy \& Environmental Science} \textbf{2020}, \emph{13}, 10
  3489.

\bibitem{Jyothirmai:2019gb}
M.~V. Jyothirmai, H.~Saini, N.~Park, R.~Thapa,
\newblock \emph{Scientific reports} \textbf{2019}, \emph{9}, 1 1.

\bibitem{li2018cation}
J.~Li, D.~Wang, X.~Li, Y.~Zeng, Y.~Zhang,
\newblock \emph{Advanced Science} \textbf{2018}, \emph{5}, 4 1700744.

\bibitem{neuschitzer2018revealing}
M.~Neuschitzer, M.~E. Rodriguez, M.~Guc, J.~A. Marquez, S.~Giraldo, I.~Forbes,
  A.~Perez-Rodriguez, E.~Saucedo,
\newblock \emph{Journal of Materials Chemistry A} \textbf{2018}, \emph{6}, 25
  11759.

\bibitem{yuan2015engineering}
Z.-K. Yuan, S.~Chen, H.~Xiang, X.-G. Gong, A.~Walsh, J.-S. Park, I.~Repins,
  S.-H. Wei,
\newblock \emph{Advanced Functional Materials} \textbf{2015}, \emph{25}, 43
  6733.

\bibitem{sharif2020control}
M.~H. Sharif, T.~Enkhbat, E.~Enkhbayar, J.~Kim,
\newblock \emph{ACS Applied Energy Materials} \textbf{2020}, \emph{3}, 9 8500.

\bibitem{khelifi2021path}
S.~Khelifi, G.~Brammertz, L.~Choubrac, M.~Batuk, S.~Yang, M.~Meuris,
  N.~Barreau, J.~Hadermann, H.~Vrielinck, D.~Poelman, et~al.,
\newblock \emph{Solar Energy Materials and Solar Cells} \textbf{2021},
  \emph{219} 110824.

\bibitem{vermang2019wide}
B.~Vermang, G.~Brammertz, M.~Meuris, T.~Schnabel, E.~Ahlswede, L.~Choubrac,
  S.~Harel, C.~Cardinaud, L.~Arzel, N.~Barreau, et~al.,
\newblock \emph{Sustainable Energy \& Fuels} \textbf{2019}, \emph{3}, 9 2246.

\bibitem{choubrac2020sn}
L.~Choubrac, M.~B{\"a}r, X.~Kozina, R.~Felix, R.~G. Wilks, G.~Brammertz,
  S.~Levcenko, L.~Arzel, N.~Barreau, S.~Harel, et~al.,
\newblock \emph{ACS Applied Energy Materials} \textbf{2020}, \emph{3}, 6 5830.

\bibitem{giraldo2015large}
S.~Giraldo, M.~Neuschitzer, T.~Thersleff, S.~L{\'o}pez-Marino, Y.~S{\'a}nchez,
  H.~Xie, M.~Colina, M.~Placidi, P.~Pistor, V.~Izquierdo-Roca, et~al.,
\newblock \emph{Advanced Energy Materials} \textbf{2015}, \emph{5}, 21 1501070.

\bibitem{ratz2019physical}
T.~Ratz, G.~Brammertz, R.~Caballero, M.~Le{\'o}n, S.~Canulescu, J.~Schou,
  L.~G{\"u}tay, D.~Pareek, T.~Taskesen, D.-H. Kim, et~al.,
\newblock \emph{Journal of Physics: Energy} \textbf{2019}, \emph{1}, 4 042003.

\bibitem{du2021defect}
Y.~Du, S.~Wang, Q.~Tian, Y.~Zhao, X.~Chang, H.~Xiao, Y.~Deng, S.~Chen, S.~Wu,
  S.~Liu,
\newblock \emph{Advanced Functional Materials} \textbf{2021}, \emph{31}, 16
  2010325.

\bibitem{kim2016improvement}
S.~Kim, K.~M. Kim, H.~Tampo, H.~Shibata, S.~Niki,
\newblock \emph{Applied Physics Express} \textbf{2016}, \emph{9}, 10 102301.

\bibitem{giraldo2018small}
S.~Giraldo, E.~Saucedo, M.~Neuschitzer, F.~Oliva, M.~Placidi, X.~Alcob{\'e},
  V.~Izquierdo-Roca, S.~Kim, H.~Tampo, H.~Shibata, et~al.,
\newblock \emph{Energy \& Environmental Science} \textbf{2018}, \emph{11}, 3
  582.

\bibitem{deng2021adjusting}
Y.~Deng, Z.~Zhou, X.~Zhang, L.~Cao, W.~Zhou, D.~Kou, Y.~Qi, S.~Yuan, Z.~Zheng,
  S.~Wu,
\newblock \emph{Journal of Energy Chemistry} \textbf{2021}, \emph{61} 1.

\bibitem{fu2020ag}
J.~Fu, D.~Kou, W.~Zhou, Z.~Zhou, S.~Yuan, Y.~Qi, S.~Wu,
\newblock \emph{Journal of Materials Chemistry A} \textbf{2020}, \emph{8}, 42
  22292.

\bibitem{he2021systematic}
M.~He, J.~Huang, J.~Li, J.~S. Jang, U.~P. Suryawanshi, C.~Yan, K.~Sun, J.~Cong,
  Y.~Zhang, H.~Kampwerth, et~al.,
\newblock \emph{Advanced Functional Materials} \textbf{2021}, 2104528.

\bibitem{wang2014device}
W.~Wang, M.~T. Winkler, O.~Gunawan, T.~Gokmen, T.~K. Todorov, Y.~Zhu, D.~B.
  Mitzi,
\newblock \emph{Advanced energy materials} \textbf{2014}, \emph{4}, 7 1301465.

\bibitem{ratz2021opto}
T.~Ratz, J.-Y. Raty, G.~Brammertz, B.~Vermang, N.~D. Nguyen,
\newblock \emph{Journal of Physics: Energy} \textbf{2021}, \emph{3}, 3 035005.

\bibitem{nishihara2017first}
H.~Nishihara, T.~Maeda, A.~Shigemi, T.~Wada,
\newblock \emph{Japanese Journal of Applied Physics} \textbf{2017}, \emph{56},
  4S 04CS08.

\bibitem{li2019effective}
J.~Li, Z.-K. Yuan, S.~Chen, X.-G. Gong, S.-H. Wei,
\newblock \emph{Chemistry of Materials} \textbf{2019}, \emph{31}, 3 826.

\bibitem{freysoldt2014first}
C.~Freysoldt, B.~Grabowski, T.~Hickel, J.~Neugebauer, G.~Kresse, A.~Janotti,
  C.~G. Van~de Walle,
\newblock \emph{Reviews of modern physics} \textbf{2014}, \emph{86}, 1 253.

\bibitem{lany2008assessment}
S.~Lany, A.~Zunger,
\newblock \emph{Physical Review B} \textbf{2008}, \emph{78}, 23 235104.

\bibitem{persson2005n}
C.~Persson, Y.-J. Zhao, S.~Lany, A.~Zunger,
\newblock \emph{Physical Review B} \textbf{2005}, \emph{72}, 3 035211.

\bibitem{park2018point}
J.~S. Park, S.~Kim, Z.~Xie, A.~Walsh,
\newblock \emph{Nature Reviews Materials} \textbf{2018}, \emph{3}, 7 194.

\bibitem{just2016secondary}
J.~Just, C.~M. Sutter-Fella, D.~L{\"u}tzenkirchen-Hecht, R.~Frahm, S.~Schorr,
  T.~Unold,
\newblock \emph{Physical Chemistry Chemical Physics} \textbf{2016}, \emph{18},
  23 15988.

\bibitem{Wang:2013gs}
W.~Wang, M.~T. Winkler, O.~Gunawan, T.~Gokmen, T.~K. Todorov, Y.~Zhu, D.~B.
  Mitzi,
\newblock \emph{Advanced Energy Materials} \textbf{2013}, \emph{4}, 7 1301465.

\bibitem{slater1964atomic}
J.~C. Slater,
\newblock \emph{The Journal of Chemical Physics} \textbf{1964}, \emph{41}, 10
  3199.

\bibitem{ma2019origin}
S.~Ma, H.~Li, J.~Hong, H.~Wang, X.~Lu, Y.~Chen, L.~Sun, F.~Yue, J.~W. Tomm,
  J.~Chu, et~al.,
\newblock \emph{The journal of physical chemistry letters} \textbf{2019},
  \emph{10}, 24 7929.

\bibitem{rey2018origin}
G.~Rey, G.~Larramona, S.~Bourdais, C.~Chon{\'e}, B.~Delatouche, A.~Jacob,
  G.~Dennler, S.~Siebentritt,
\newblock \emph{Solar Energy Materials and Solar Cells} \textbf{2018},
  \emph{179} 142.

\bibitem{alkauskas2014first}
A.~Alkauskas, Q.~Yan, C.~G. Van~de Walle,
\newblock \emph{Physical Review B} \textbf{2014}, \emph{90}, 7 075202.

\bibitem{kresse1996efficiency}
G.~Kresse, J.~Furthm{\"u}ller,
\newblock \emph{Computational Materials Science} \textbf{1996}, \emph{6}, 1 15.

\bibitem{kresse1999ultrasoft}
G.~Kresse, D.~Joubert,
\newblock \emph{Physical Review B} \textbf{1999}, \emph{59}, 3 1758.

\bibitem{perdew1996generalized}
J.~P. Perdew, K.~Burke, M.~Ernzerhof,
\newblock \emph{Physical Review Letters} \textbf{1996}, \emph{77}, 18 3865.

\bibitem{sun2015strongly}
J.~Sun, A.~Ruzsinszky, J.~P. Perdew,
\newblock \emph{Physical Review Letters} \textbf{2015}, \emph{115}, 3 036402.

\bibitem{sun2016accurate}
J.~Sun, R.~C. Remsing, Y.~Zhang, Z.~Sun, A.~Ruzsinszky, H.~Peng, Z.~Yang,
  A.~Paul, U.~Waghmare, X.~Wu, et~al.,
\newblock \emph{Nature Chemistry} \textbf{2016}, \emph{8}, 9 831.

\bibitem{heyd2003hybrid}
J.~Heyd, G.~E. Scuseria, M.~Ernzerhof,
\newblock \emph{The Journal of Chemical Physics} \textbf{2003}, \emph{118}, 18
  8207.

\bibitem{fritsch2020climbing}
D.~Fritsch, S.~Schorr,
\newblock \emph{Journal of Physics: Energy} \textbf{2020}, \emph{3}, 1 015002.

\end{thebibliography}

\end{document}